\documentclass[epj]{svjour}
\usepackage{amssymb}
\usepackage{graphicx}
\usepackage{dcolumn,color}
\usepackage{bm}
\usepackage{epstopdf}
\usepackage{epsfig,subfigure}
\usepackage{color,framed}
\usepackage{amsmath,amssymb}
\begin{document}
\title{Localization of $q$-form field on squared curvature gravity domain wall brane coupling with gravity and background scalar}

\author{Xin-Nuo Zhang, Heng Guo \and Yong-Tao Lu}                     
%
\mail{hguo@xidian.edu.cn}
\institute{School of Physics, Xidian University, Xi'an 710071, China}
\date{Received: date / Revised version: date}
%
\abstract{Unlike the duality in four-dimensional spacetime, where the $q-$form fields corresponds to the scalar fields
or the vector fields, in higher dimensional spacetime, they denote new types of particles. In this paper, we investigate
the localization of the KK modes of the $q$-form fields in a five dimensional brane world. We introduce the coupling
between the $q-$form fields and both the gravity and the background scalar field. This consideration enables the
localization of the $q$-form fields on the five-dimensional RS-2 thick brane, leading to the derivation of zero modes,
Schr\"{o}dinger-like equations, and a four-dimensional effective action. We suggest a new function $F(R,\varphi)$ to
stand for the coupling of the $q$-form field with gravity and background scalar fields. Our analysis highlights the
significance of the parameters $\text C_2$ and $\displaystyle t$ in the localization processes.
\PACS{
      ***
     }
} 
\maketitle
\section{Introduction}
\label{sec:intro}
For nearly a century, the concept of extra dimensions has been a tantalizing topic in theoretical physics. In the early 20th
century, Kaluza and Klein pioneered the exploration of extra dimensions with their formulation of a five-dimensional spacetime
theory, famously known as the Kaluza-Klein (KK) theory \cite{8,9}, aiming to unify electromagnetic and gravitational forces.
This groundbreaking idea paved the way for further investigations into the physics of extra dimensions. However in this scenario,
the size of the extra dimension is so extremely small that it cannot be detected by the present experiments. Entering the 21st
century, after the Arkani-Hamed-Dimopoulos-Dvali (ADD) model \cite{89} was put forward, Randall and Sundrum introduced models
with warped extra dimensions, known as the RS-1 and RS-2 models \cite{10,11}, to address the long-standing hierarchy problem
and the cosmological constant problem.

{
However, in the above models \cite{89,10,11}, the brane is idealized as having no thickness, which leads to a singularity
at the brane location. In the most fundamental theory, there appears to exist a minimum length scale, and thus the 
thickness of a brane should be taken into account in more realistic and physically meaningful field models. For this 
reason, more natural thick brane scenarios have been extensively investigated
\cite{12,13,14,Wang0201051,Bazeia0610233,Afonso0601069,Slatyer0609003,LYX1104.3188,Bazeia1306.1847,ZY1507.00630,WJJ2010.05016,22,22-1}.
Actually in the early 1980s, the domain wall model which derives from the topology defect, was proposed as a thick 
brane without considering the gravity \cite{RubakovPLB1983}. Subsequently, over the last three decades, based on gravity 
coupled to the scalars, many smooth thick branes have been naturally constructed \cite{12,13,14,Wang0201051,Bazeia0610233,Afonso0601069,Slatyer0609003,WJJ2010.05016}.
For more details, see refs. \cite{VDzhunushaliev0904.1775,6}.

In addition to thick branes generated by scalar fields, one can also consider the $f(R)$ theory, in which
the Lagrangian is given by a function of the scalar curvature $R$. This theory was originally
proposed in cosmology, and is widely regarded in literature as a valid alternative to Dark Energy, Dark
Matter and for the possibility of generating inflation (e.g. the Starobinsky model)
\cite{Capozziello0201033,Alexei0706.2041,Sahni0610026,Capozziello1108.6266,Sotiriou0805.1726,DeFelice1002.4928,Nojiri1011.0544}.
Reference \cite{Capozziello0201033} achieved exact accelerated expanding solutions for several fourth order
theory of gravity, and got an alternative scheme to the standard quintessence scalar field. For higher-derivative
$f(R)$ gravity, a class of models was also proposed which produce viable cosmology different from the LambdaCDM
(Lambda Cold Dark Matter) one \cite{Alexei0706.2041}. 

Within the braneworld framework, a number of works have been devoted to embedding branes into various types of 
$f(R)$ theories \cite{LYX1104.3188,Bazeia1306.1847,ZY1507.00630,22,22-1,6}.
In refs. \cite{LYX1104.3188,Bazeia1306.1847}, thick $f(R)$ branes are generated by  background scalar
fields, while in refs. \cite{ZY1507.00630,22,22-1}, they are derived from the pure geometric $f(R)$ gravity
without introducing additional matter fields. Besides, other approaches to constructing thick brane models 
have also been explored (see refs. \cite{VDzhunushaliev0904.1775,6} and references therein).

Despite significant progress in understanding extra dimensions \cite{22,23,24,25,26,27,28,29,30,31,32,33,34,35,36,37,38,39,40,41,42,43,44,45,46},
there remains a theoretical gap in the study of the localization of various matter fields on the domain wall
branes, such as the localization of the $q$-form fields \cite{6,1,3,4,5,7,2}. Specifically, the effective
localization of the five-dimensional (5D) $q$-form fields on four-dimensional (4D) spacetime \cite{3,4}, and
the implications of such localization for theoretical physics and cosmology, pose pressing questions that need
to be addressed. Motivated by this gap, this paper considers a coupling mechanism between the $q$-form fields
and both the gravity and the background scalar field, and delves into the localization of the $q$-form fields
on the brane. One of the most promising modified gravity theories is non-local gravity, in which the action is
expressed as a function of the scalar curvature $R$
\cite{SNojiri9706009,Elizalde9508041,Capozziello2107.06972,SNojiri1911.07329,Capozziello2201.04512},
with similar functional forms also appearing in gravity coupling case \cite{44,5,2}. On the other hand, for
relativistic particles, the kinetic energy is much larger than the rest energy. Therefore, it is both reasonable
and physically motivated to consider this coupling mechanism. Furthermore, this paper successfully addresses the
localization of the 0-form, 1-form, and 2-form fields on an RS-like thick brane, examining the behaviors of the
KK modes of these fields when individually coupled with gravity and the background scalar field.

In this scenario, there is a coupling factor \(F(R,\varphi)\) in the action of the $q-$form fields. When coupled
with the gravity, the KK modes of various $q-$form fields will exhibit same localization properties. The zero-mode
can always be localized on the brane. Then, there is a critical value $1/2$ for the coupling parameter $\text C_2$. If
$0<\text C_2<1/2$, there is a volcanic potential, and no localized massive mode. If $\text C_2=1/2$, a finitely
deep potential well emerges, and a finite number of massive modes can be localized with the number rising alongside
the parameter $\text C_1$. Lastly if $\text C_2>1/2$, there will be an infinitely deep well and an infinite discrete
mass spectrum. When coupled with the background scalar field, the effective potential of various $q-$form fields
are the same form of volcano potential. However, the localization of the scalar field imposes no requirement on the
coupling parameter $t$, while for the $U(1)$ gauge vector field, it requires $t>1/2$, and for the Kalb-Ramond (KR)
field, it requires $t>3/2$.


The remaining parts are organized as follows. In section \ref{Sec22}, we present a specific RS thick
brane-world solution. Next, a concise overview of our methodology for localization of $q$-form field
is introduced in section \ref{Sec33}. Then in section \ref{Sec44}, we discuss the localization of
various $q$-form fields-specifically, the scalar fields, the $U(1)$ gauge vector fields, and the KR
fields-on the domain wall brane. Finally, the conclusions are given in section \ref{Sec55}.
}

\section{Review of squared curvature gravity} \label{Sec22}
In this section, we briefly revisit the domain wall brane theory in the context of curvature gravity \cite{1}. This
theory is endowed with the following action:
\begin{equation} \label{eq1}
S_{\text{brane}}=\int{d^5x\sqrt{-g}\left( \frac{1}{2\kappa _{5}^{2}}f\left( R \right)
                 -\frac{1}{2}\partial ^M\varphi \partial _M\varphi -V\left( \varphi \right) \right)},
\end{equation}
where $\kappa _{5}^{2}=8\pi G_5$ is the five-dimensional gravitational coupling constant, with $G_5$ being the
five-dimensional Newton constant. For simplicity, we set $\kappa_5^{2} = 1$. The factor $f\left(R\right)$ is a
function of the scalar curvature $R$. Additionally, $V\left( \varphi \right)$ is the background potential with
$\varphi$ the background scalar field.

In the case of flat brane, the line element is given by:
\begin{equation} \label{eq2}
ds^2=g_{MN}dx^Mdx^N=e^{2A\left( y \right)}\eta _{\mu \nu}dx^{\mu}dx^{\nu}+dy^2.
\end{equation}
Here, $e^{2A\left(y\right)}$ is the warp factor, and $\eta_{\mu\nu}=\text{diag}\left(-,+,+,+\right)$ is the
metric of the brane. The capital Latin letters $\displaystyle M,N=0,1,2,3,5$ and the Greek letters
$\displaystyle \mu,\nu=0,1,2,3$ are uesd to represent the bulk and brane indices, respectively. Moreover,
the scalar field $\varphi =\varphi \left( y \right) $ depends only on the extra-dimensional coordinate.

Based on the action (\ref{eq1}), the Einstein equations can be obtained:
\begin{eqnarray}
f\left( R \right) +2f_R\left( 4A'^2+A'' \right)\hspace{-0.15cm}&-&\hspace{-0.15cm}6f_{R}'A'-2f_{R}''       \nonumber    \\
                 & &\hspace{-0.15cm}=\kappa _{5}^{2}\left(\varphi ^2+2V\left(\varphi \right)\right),       \label{eq3}         \\
-8f_R\left( A''+A'^2 \right) +8f_{R}'A'\hspace{-0.15cm}&-&\hspace{-0.15cm}f\left( R \right)                                \nonumber    \\
                 & &\hspace{-0.15cm}=\kappa _{5}^{2}\left( \varphi '^2-2V\left(\varphi\right) \right),     \label{eq4}
\end{eqnarray}
where $f_R=\frac{df\left( R \right)}{dR}$, and the prime denotes the derivatives with respect to the coordinate $y$.

The background scalar field $\varphi$ is described by the following equation:
\begin{equation} \label{eq5}
4A'\varphi '+\varphi ''-\frac{\partial V}{\partial \varphi}=0.
\end{equation}
We consider the function $f(R)$ of the form
\begin{equation} \label{eq6}
f\left( R \right) =R+\alpha R^{2},
\end{equation}
and then $f_R=1+2\alpha R$, $f_{R}'=2\alpha R'$, $f_{R}''=2\alpha R''$. When $\alpha =0$, it reverts to
the case of General Relativity, and for $\alpha \ne 0$, the Einstein equations become the fourth-order
differential ones.

To find a brane world solution, the potential $V(\varphi)$ can be considered as:
\begin{equation} \label{eq7}
V\left( \varphi \right) =\lambda ^{\left( 5 \right)}\left( \varphi ^2-\upsilon ^2 \right) ^2+ \Lambda _5,
\end{equation}
where $\displaystyle \lambda ^{\left( 5 \right)}>0$ is the self-coupling constant of the scalar field, and
$\displaystyle \Lambda _5$ is the five-dimensional cosmological constant, which is responsible for the dark
energy problem and negative in AdS$_5$ spacetime.

A simple form of the solution can be obtained:
\begin{eqnarray}
   e^{A\left( y \right)} &=&\text{sech} \left( ky \right),               \label{eq9-1}     \\
   \varphi \left( y \right) &=& \upsilon \tanh \left( ky \right),        \label{eq8}
\end{eqnarray}
where
\begin{equation} \label{eq9}
\lambda ^{\left( 5 \right)}=\frac{3}{784\alpha},\, \upsilon =7\sqrt{\frac{3}{29}},\, \Lambda _5=-\frac{477}{6728\alpha},\, k=\sqrt{\frac{3}{232\alpha}},
\end{equation}
and $\alpha$ is an undetermined constant. Then, we can obtain
\begin{eqnarray}
A\left( y \right) &=& -\ln \left( \cosh \left( ky \right) \right),      \label{eq10}   \\
R\left( y \right) &=& -20k^2\tanh ^2\left( ky \right) +8k^2\text{sech} ^2\left( ky \right).      \label{eq11}
\end{eqnarray}
As $y\rightarrow\infty$, $A\left( y \right) \rightarrow -k|y|+\text C$, the bulk scalar curvature $R\rightarrow -20k^2+24k^2\text{e}^{- 2k|y|}$.
Moreover, the background scalar field  $\varphi(y)$ is a kink solution, satisfying
$\varphi \left( 0 \right) =0,\varphi \left( \pm \infty \right) =\pm \upsilon$, so the background potential $V\left( \varphi \right) $
takes its minimum at $\varphi =\pm \upsilon $.

By reviewing the aforementioned theory on the domain wall brane, we find that a straightforward thick brane
solution can be achieved with $f\left( R \right) =R+\alpha R^2$. However, subsequent studies discovered that
only the $q=0$ scalar field can be localized on the brane, the $q=1$ vector field and the $q=2$ KR field cannot
be localized on the brane due to the non-normalization of the zero mode.
To address this issue, we will attempt to introduce a new localization method for the $q$-form fields and
discuss the feasibility of this method in the next part of this article, with the computational results
presented in specific forms.

\section{Localization of $q-$form field on the brane} \label{Sec33}
With considering the $q-$form field coupling with the gravity and the background scalar field, the action of a free
$q$-form field $\displaystyle X_{M_1M_2...M_q}$ can be given by
\begin{eqnarray}\label{eq12}
S_q &=& -\frac{1}{2(q+1)!}\int d^5x \sqrt{-g}F(R,\varphi)     \nonumber \\
   & & \times Y_{M_1M_2\cdots M_{q+1}}Y^{M_1M_2\cdots M_{q+1}},
\end{eqnarray}
where the field strength $Y_{M_1M_2\cdots M_{q+1}} = \partial_{[M_1}X_{M_2\cdots M_{q+1}]}$. The function $F(R,\varphi)$
stands for the coupling of the $q$-form field with gravity and the background scalar field.

To facilitate the discussion on the localization of massive modes, it is advantageous to perform the following coordinate
transformation:
\begin{equation} \label{eq13}
\begin{cases}
dz = \text{e}^{-A(y)}dy,\\
z = \int \text{e}^{-A(y)}dy
\end{cases}
\end{equation}
with the boundary condition $z(y=0)=0$. Then, the line element (\ref{eq2}) can be expressed in terms
of the coordinate $z$ as
\begin{equation} \label{eq14}
ds^2 = \text{e}^{2A(z)}(\eta_{\mu\nu}dx^{\mu}dx^{\nu} + dz^2),
\end{equation}
where $z$ is the conformal coordinate of $y$, the factor $\text{e}^{2A(z)}$ means that this metric is non-factorizable.

With this metric (\ref{eq14}) and the gauge choice
\begin{equation} \label{eq14-1}
X_{zM_2\cdots M_q}(x_{\mu},z)=0,
\end{equation}
the equations of motion derived from the action (\ref{eq12}) are
\begin{eqnarray}
\partial_{\mu_1}(\sqrt{-g} F(R,\varphi) Y^{\mu_1\mu_2\cdots \mu_{q+1}})& &     \nonumber     \\
+ \partial_z(\sqrt{-g} F(R,\varphi) Y^{z\mu_2\cdots \mu_{q+1}}) &=& 0,       \label{eq15-1}       \\
\partial_{\mu_1}(\sqrt{-g} F(R,\varphi) Y^{\mu_1\cdots \mu_qz}) &=& 0.       \label{eq15-2}
\end{eqnarray}

To achieve the localization of the $q$-form field on the domain wall brane, we first make a gauge choice
\begin{eqnarray} \label{gauge}
X_{\mu_1\cdots \mu_{q-1}\mu_z}=0,
\end{eqnarray}
and conduct a novel method of KK decomposition:
\begin{eqnarray} \label{eq16}
X_{\mu_1\mu_2\cdots \mu_q}(x_{\mu}, z) &=& \sum_n \hat{X}_{\mu_1\mu_2\cdots \mu_q}^{(n)}(x^{\mu}) U^{(n)}(z)      \nonumber  \\
     & & \times \text{e}^{\frac{2q-3}{2} A} F(R,\varphi)^{-\frac{1}{2}}.
\end{eqnarray}
Here, $U^{(n)}(z)$ represent the KK modes of the $q$-form field. Then, the field strength becomes:
\begin{eqnarray}
Y_{\mu_1\mu_2\cdots \mu_{q+1}}(x^{\mu}, z) &=&
            \sum_n \hat{Y}_{\mu_1\mu_2\cdots \mu_{q+1}}^{(n)}(x^{\mu}) {U}^{(n)}(z)    \nonumber   \\
            & & \times \text{e}^{\frac{2q-3}{2} A} F(R,\varphi)^{-\frac{1}{2}},   \label{eq17-1}      \\
{Y}_{\mu_1\mu_2\cdots \mu_qz}(x^{\mu}, z) &=&
            \sum_n \frac{1}{q+1} \hat{X}_{\mu_1\mu_2\cdots \mu_q}^{(n)}(x^{\mu})               \nonumber \\
      \times \bigg({U}'^{(n)}\hspace{-0.1cm}&F&\hspace{-0.1cm}(R,\varphi)^{-\frac12}
         -\frac12U^{(n)}F(R,\varphi)^{-\frac32}F'(R,\varphi)                                   \nonumber \\
 + \frac{2q-3}{2}\hspace{-0.1cm}&A&\hspace{-0.1cm}'{U}^{(n)}F(R,\varphi)^{-\frac12} \bigg)\text{e}^{\frac{2q-3}{2} A}.         \label{eq17-2}
\end{eqnarray}

Substituting these relations (\ref{eq17-1}) and (\ref{eq17-2}) into the equations of motion (\ref{eq15-1})
yields a Schr\"{o}dinger-like equation for the KK modes of the $q$-form field:
\begin{equation} \label{eq18}
\left[ -\partial_z^2 + V(z) \right] {U}^{(n)}(z) = m_n^2 {U}^{(n)}(z),
\end{equation}
where $m_n$ is the mass of the KK modes, and the effective potential $V(z)$ is given by:
\begin{eqnarray} \label{eq19}
V(z) &=& \frac{(3-2q)^2}{4}A'^2(z) + \frac{3-2q}{2}A''(z) + \frac{F''(R,\varphi)}{2F(R,\varphi)}  \nonumber   \\
  & &+ \frac{(3-2q) F'(R,\varphi)}{2F(R,\varphi)}A'(z) - \frac{F'^2(R,\varphi)}{4F^2(R,\varphi)}
\end{eqnarray}
with the prime standing for the derivative with respect to $z$.

The orthogonal normalization for ${U}^{(n)}$ is a prerequisite, requiring
\begin{equation} \label{eq20}
\int {U}^{(m)} {U}^{(n)} dz = \delta_{mn},
\end{equation}
which allows us to reduce the action (\ref{eq12}) into the 4D effective one as
\begin{eqnarray} \label{eq21}
S_{\text{eff}} &=& \sum_n \int d^4x \sqrt{-\eta} \bigg( \hat{Y}_{\mu_1\mu_2\cdots \mu_{q+1}}
         \hat{Y}^{\mu_1\mu_2\cdots \mu_{q+1}}    \nonumber \\
    & &+ \frac{1}{q+1} m_n^2 \hat{X}_{\mu_1\mu_2\cdots \mu_q}^{(n)} \hat{X}^{(n)\mu_1\mu_2\cdots \mu_q} \bigg).
\end{eqnarray}
By satisfying the orthogonal normalization, the KK modes of the $q$-form field can be localized on the brane.

In addition, the zero-mode solution $U^{(0)}(z)$ can be obtained by factorizing the Schr\"{o}dinger-like
equation (\ref{eq18}). The transformations are as follows:
\begin{equation} \label{eq22}
QQ^{\dagger}{U}^{(n)}(z) = m_{n}^{2}{U}^{(n)}(z),
\end{equation}
where
\begin{eqnarray}
Q &=& \partial_{z} +\left( \frac{3-2q}{2}A'+ \frac{F'(R,\varphi)}{2F(R,\varphi)} \right),      \label{eq23}    \\
Q^{\dagger} &=& -\partial_{z} + \left( \frac{3-2q}{2}A'+ \frac{F'(R,\varphi)}{2F(R,\varphi)} \right).    \label{eq24}
\end{eqnarray}
Equation (\ref{eq22}) implies that the square mass $m_n^{2}$ is non-negative, ensuring that no tachyonic field \cite{20,21}
forms. By setting $m_{0}^{2} = 0$, the zero-mode solution can be obtained as:
\begin{equation} \label{eq25}
{U}^{(0)}(z) = N\text{e}^{\frac{3-2q}{2}A}F(R,\varphi)^{\frac{1}{2}},
\end{equation}
where $N$ is the normalization constant. A normalizable zero-mode means that the KK mode with $m_0^2 = 0$ can be localized,
and exactly reproduces the particle on the brane.

{
Finally, we specifically discuss the form and properties of the coupling function $F( R,\varphi )$. This function
depends on both the scalar curvature $R$ and the background scalar $\varphi$. Naturally, in the decoupling case with
$F\left( R,\varphi \right)=1$, the action (\ref{eq12}) reverts to the standard form one. Similar functions have been
discussed in refs. \cite{SNojiri9706009,Elizalde9508041,Capozziello2107.06972,SNojiri1911.07329,Capozziello2201.04512}.
For clarity, we will separately investigate the coupling of the $q-$form fields with the gravity and with
the background scalar field in the following.

$Case$ I: Coupling with the gravity, $F\left(R,\varphi\right)=G\left(R\right)$. The function $G\left( R\right )$
should adhere to the following criteria:
\begin{enumerate}
    \item The 5D scalar curvature $R$ and the coupling function $G(R)$ should be nonsingular.
    \item In a region without gravity, the 5D spacetime becomes flat and its scalar curvature $R\rightarrow0$, so the coupling
          turns into the minimal coupling with $G(R)\rightarrow1$, and the action returns to the one of a free field.
    \item The function $G(R)$ should satisfy the positivity condition
    \begin{equation}
      G(R) >0 \label{positivity}
    \end{equation}
    to preserve the canonical form of 4D action.
\end{enumerate}

Concerning the specific form of function $G(R)$, the polynomial of $R$ are employed in studying the
non-local gravity \cite{Capozziello2201.04512}. In addition, for the localization of the matter fields \cite{5,2},
both polynomial and exponential functions of $R$ have been considered. However, the exponential function can always
be expanded as a polynomial. Moreover, the dilaton scalar field $\pi$ typically emerges in the exponential form
$e^{\xi\pi}$ when coupled to matter fields \cite{27,FuCE1101.0336}. This dilaton field is of even-parity, the same
as the scalar curvature $R$ in the brane model (\ref{eq9-1}). Therefore, we suggest the coupling function
$G\left( R\right )$ takes the form as
\begin{equation} \label{eq26}
G\left( R \right) =\text{e}^{\text C_1\left( 1-\left( 1+\frac{R}{20k^2} \right) ^{-\text C_2} \right)},
\end{equation}
where $\text C_1$ and $\text C_2$ are positive parameters. The parameter $\text C_2$ is closely related to the
effective potentials of the KK modes in the corresponding Schr\"{o}dinger equations.

$Case$ II: Coupling with the background scalar field, $F\left(R,\varphi\right)=\tilde F\left(\varphi\right)$. As a
factor in the action (\ref{eq12}), function $\tilde F\left(\varphi\right)$ should always be positive. In light of
the background scalar (\ref{eq8}), we consider the coupling function $\tilde F\left(\varphi\right)$ of the form
\begin{equation} \label{eqFphi}
\tilde F\left(\varphi\right)=\lambda\left(\varphi_0^2-\varphi ^2\right)^t,
\end{equation}
where $\lambda$ is the normalization constant, $\varphi_0$ is a undefined parameter, and $t$ is a positive
coupling parameter. For this coupling, it should return to the minimal coupling with $\tilde F(\varphi)=1$
when the background scalar vanishes. So, we can get the normalization constant
\begin{equation} \label{NormLamd}
\lambda=\varphi_0^{-2t}.
\end{equation}
Furthermore, using the coordinate transformation (\ref{eq13}), the background scalar field $\varphi$ can be
expressed in terms of coordinate $z$ as
\begin{equation} \label{eq27}
\varphi \left( z \right) =\pm 7\sqrt{\frac{3}{29}}\sqrt{1-\frac{1}{1+k^2z^2}},
\end{equation}
and then the parameter $\varphi_0$ is chosen as:
\begin{equation} \label{eq28}
\varphi _0=|\varphi \left( \pm \infty \right)| = 7\sqrt{\frac{3}{29}}.
\end{equation}
}

Based on the above decomposition and transformation, we have successfully obtained the zero-mode solution (\ref{eq25})
of the $q$-form field. Furthermore, by requiring the orthogonal normalization condition (\ref{eq20}), we can deduce
the 4D effective action (\ref{eq21}). At last, the core of the problem is focused on the effective potential (\ref{eq19}),
which is decided by the warp factor $A(z)$, the coupling function $F(R,\varphi)$, and the index $q$. This is just the
clue of the next section.

\section{Localization of various $q-$form fields} \label{Sec44}

In this section, we explore the localization of various $q$-form fields within the 5D RS-like brane model.
Certainly, it is implicitly assumed that various $q-$form fields considered here make little contribution
to the bulk energy, so that the brane solution given below remains valid even in the presence of bulk fields.

Applying the transformation (\ref{eq13}) to Eq. (\ref{eq10}), we derive the specific expression of $A(z)$ as follows:
\begin{equation} \label{eq29}
A\left( z \right) = -\frac{1}{2}\log\left( 1+k^2z^2 \right).
\end{equation}
Then, the scalar curvature $R(z)$ can be expressed as
\begin{eqnarray} \label{eq30}
R(z) &=& 4e^{-2A(z)}\left( -2\partial_{z,z}A(z) -3\partial_{z}^{2}A(z) \right)      \nonumber   \\
     &=& \frac{8k^2 - 20k^4z^2}{1+k^2z^2}.
\end{eqnarray}
The asymptotic behavior of $R(z)$ is
\begin{equation} \label{eq31}
R\left( z\rightarrow \infty \right) \rightarrow -20k^2 + \frac{8}{z^2}.
\end{equation}
Upon examining the asymptotic behavior of $R(z)$, it can be observed that, both in the coordinate $y$
and in the coordinate $z$, $R(z)$ converges to the same finite value at infinity of the extra dimension.

Building on the Schr\"{o}dinger-like equation (\ref{eq18}) and the zero-mode solution (\ref{eq25}),
we proceed to specify their behaviors for different field scenarios with specific brane model.

\subsection{Scalar field}
In the context of quantum field theory, the $q=0$ scalar fields, characterized by spin 0, are typically
divided into real and complex scalar fields. The process and results of localizing 5D real scalar fields
closely mirror those for the complex scalar fields. Therefore, this article will concentrate on the former.

Considering the coupling factor $F(R,\varphi)$, we suggest the a 5D scalar field of the form
\begin{equation} \label{eq32}
S_{\text{scalar}} = -\frac{1}{2}\int d^5x \sqrt{-g} F(R,\varphi) \partial_M \varPhi \partial^M \varPhi.
\end{equation}
In terms of the metric (\ref{eq14}), the equation of motion (EOM) can be derived from the action (\ref{eq32})
as:
\begin{equation} \label{eq33}
\partial_{\mu} \left( \sqrt{-g} F(R,\varphi) \partial^{\mu} \varPhi \right)
      + \partial_z \left( \sqrt{-g} F(R,\varphi) \partial^z \varPhi \right) = 0.
\end{equation}
By employing the KK decomposition
\begin{equation} \label{eq34}
\varPhi(x^{\mu}, z) = \sum_n \phi_n(x^{\mu}) \chi_n(z) e^{-\frac{3}{2}A} F(R,\varphi)^{-\frac{1}{2}},
\end{equation}
and ensuring that $\phi_n$ satisfies the Klein-Gordon equation:
\begin{equation} \label{eq35}
\left[ \frac{1}{\sqrt{-\eta}} \partial_{\mu} \left( \sqrt{-\eta} \eta^{\mu \nu} \partial_{\nu} \right) - m_{n}^{2} \right] \phi_n(x) = 0,
\end{equation}
we can obtain the following specific form of the Schr\"{o}dinger-like equation for the scalar KK modes:
\begin{equation} \label{eq36}
\left[ -\partial_{z}^{2} + V_0(z) \right] \chi_n(z) = m_{n}^{2} \chi_n(z),
\end{equation}
where the effective potential is
\begin{eqnarray} \label{eq37}
V_0(z) &=& \frac{3}{2}A''(z) + \frac94A'^2(z) + \frac{F''(R,\varphi)}{2F(R,\varphi)}      \nonumber   \\
       & & +\frac{3F'(R,\varphi)}{2F(R,\varphi)}A'(z) - \frac{F'^2(R,\varphi)}{4F^2(R,\varphi)}.
\end{eqnarray}

With requiring the orthogonal normalization:
\begin{equation} \label{eq38}
\int \chi_m(z) \chi_n(z) dz = \delta_{mn},
\end{equation}
the 5D action (\ref{eq32}) can be reduced to the 4D effective one:
\begin{equation} \label{eq39}
S_{\text{scalar}} = -\frac{1}{2} \sum_n \int d^4x \sqrt{-\eta}
                   \left( \partial_{\mu} \phi_n \partial^{\mu} \phi_n + m_{n}^{2} \phi_{n}^{2} \right).
\end{equation}
For the case of minimal coupling where $F(R,\varphi) \equiv 1$, the effective potential is
\begin{equation} \label{eq40}
V_0(z) = \frac{3}{2}A'' + \frac{9}{4}A'^2.
\end{equation}
This potential $V_0(z)$ exhibits a volcano shape, which allows only the scalar zero mode to be localized
on the brane, akin to that of the thin brane scenario. Following the method shown in Ref. \cite{7}, we
obtain the zero-mode solution for the scalar field:
\begin{equation} \label{eq41}
\chi_{0}(z) = N_0 \text{e}^{\frac{3}{2} A(z)} F(R,\varphi)^{\frac{1}{2}},
\end{equation}
where $N_0$ denotes the normalization constant.

$Case$ I:
Coupling with the gravity, $\displaystyle F(R,\varphi) = G(R)$. Firstly for the scalar zero-mode, by substituting
Eqs. (\ref{eq29}) and (\ref{eq30}) into the zero-mode solution (\ref{eq41}), we can obtain
\begin{equation} \label{eq42}
\chi_0(z) =N_0\frac{\sqrt{\text{e}^{\text C_1\left( 1-(\frac{5}{7})^{\text C_2}\!\:\!\:(\frac{1}{1+k^2\!\:z^2})^{-\text C_2}
            \right)}}}{(1+k^2\!\:z^2)^{\frac{3}{4}}}.
\end{equation}
Furthermore, if $z\rightarrow\infty$, the scalar zero mode exhibits the following asymptotic behavior:
\begin{equation} \label{eq48}
\chi_0( z\rightarrow\infty ) \rightarrow N_0k^{-\frac32}e^{\frac12\text C_1}z^{-\frac32}
                                         e^{-\frac12(\frac57)^{\text C_2}\text C_1k^{2\text C_2}z^{2\text C_2}}.
\end{equation}
From this expression, it can be seen that as parameters $\text C_1$ and $\text C_2$ are positive, the scalar
zero-mode $\chi_0$ will converge to zero exponentially when far away from the brane. Therefore, this zero-mode
is normalizable, and can be localized on the brane.

Then for the localization of the massive KK modes, it is dependent on the behaviors of the effective potential.
In light of the warp factor $A(z)$ (\ref{eq29}) and the coupling function $G(R)$ (\ref{eq26}), the effective
potential can be given by
\begin{eqnarray} \label{eq44}
V_0\left( z \right) \hspace{-0.1cm}&=&\hspace{-0.1cm}\frac{1}{4}\!\:k^2\!\:\bigg(\frac{1}{1+k^2\!\:z^2}\bigg)^{2-2\!\:\text C_2}\!\:
        \bigg[4\!\:(\frac{25}{49})^{\text C_2}\!\:{\text C_1}^2\!\:{\text C_2}^2\!\:k^2\!\:z^2       \nonumber  \\
& &\hspace{-0.1cm}+3\!\:(\frac{1}{1+k^2\!\:z^2})^{2\!\:\text C_2}\!\:(-2+5\!\:k^2\!\:z^2)-4\times 5^{\text C_2}\text C_1\!\:\text C_2    \nonumber   \\
& &\hspace{-0.1cm}\times\!\:(\frac{1}{7+7\!\:k^2\!\:z^2})^{\text C_2}\!\:(1+2\!\:(-2+\text C_2)\!\:k^2\!\:z^2)\bigg].
\end{eqnarray}
For the analysis of the specific behavior of the effective potential at infinity, we find:
\begin{equation} \label{eq46}
\begin{aligned}
V_0( z\rightarrow \infty)\rightarrow \left(\frac{25}{49}\right)^{\text C_2}\text C_1^2\text C_2^2k^{4\text C_2}z^{4\text C_2-2}.
\end{aligned}
\end{equation}
From the analysis above, we further deduce the conditions for the value of $V_0(z)$ as $z\rightarrow \infty$,
specifically:
\begin{equation} \label{eq47}
V_0\left( z\rightarrow \infty \right) \rightarrow\left\{
\begin{array}{llcll}
+\infty,         & &\text C_2  > 1/2   \\
V_{\text{const}},        & &\text C_2 = 1/2  \\
0,                 & &0<\text C_2 <1/2,
\end{array} \right.
\end{equation}
where $\displaystyle V_{\text{const}}=\frac{5}{28}\text C_1^2k^2$, and it is clear that $V_{\text{const}} > 0$.
This demonstrates that if $\text C_2=1/2$, the effective potential is a finitely deep potential well, and its
depth depends on the parameters $\text C_1$ and $k$.

\begin{figure}[htbp]
\subfigure[$V_0(z)$]{\label{fig:subfig1a}
\includegraphics[width=0.95\columnwidth]{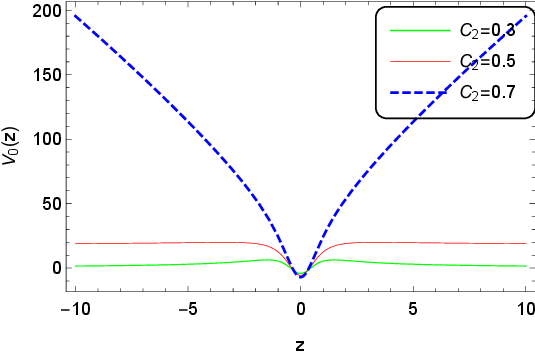}}
\subfigure[$\chi_0\left(z\right)$]{\label{fig:subfig1b}
\includegraphics[width=0.95\columnwidth]{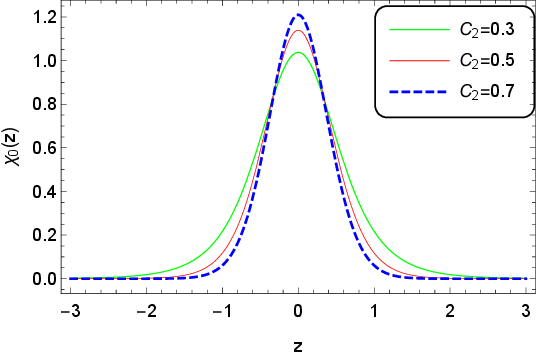}}
\caption{The effective potentials $\displaystyle V_0\left(z\right)$ in (a), and the shapes of the scalar zero-mode
         $\displaystyle \chi_0\left(z\right)$ in (b). The parameters are set as $\text C_1=10,k=1$, and
         $\text C_2=0.3,0.5,0.7$.}
\label{fig:subfig_1}
\end{figure}

In order to more intuitively reflect the content of the above analysis, we have plotted the effective potential
$V_0(z)$ and the scalar zero-mode $\chi_0(z)$ in Fig. \ref{fig:subfig_1} via numerical calculations. The figure
\ref{fig:subfig1a} demonstrates that the behaviors of the effective potential align well with the conclusions
(\ref{eq47}). Subsequently, we will explicitly outline the localization of the massive KK modes in the cases
of different values of parameters $\text C_1$ and $\text C_2$.

\begin{figure*}[htbp]
\subfigure[$\text C_1=15,k=1,\text C_2=0.5$]{\label{fig:subfig2a}
\includegraphics[width=0.31\textwidth]{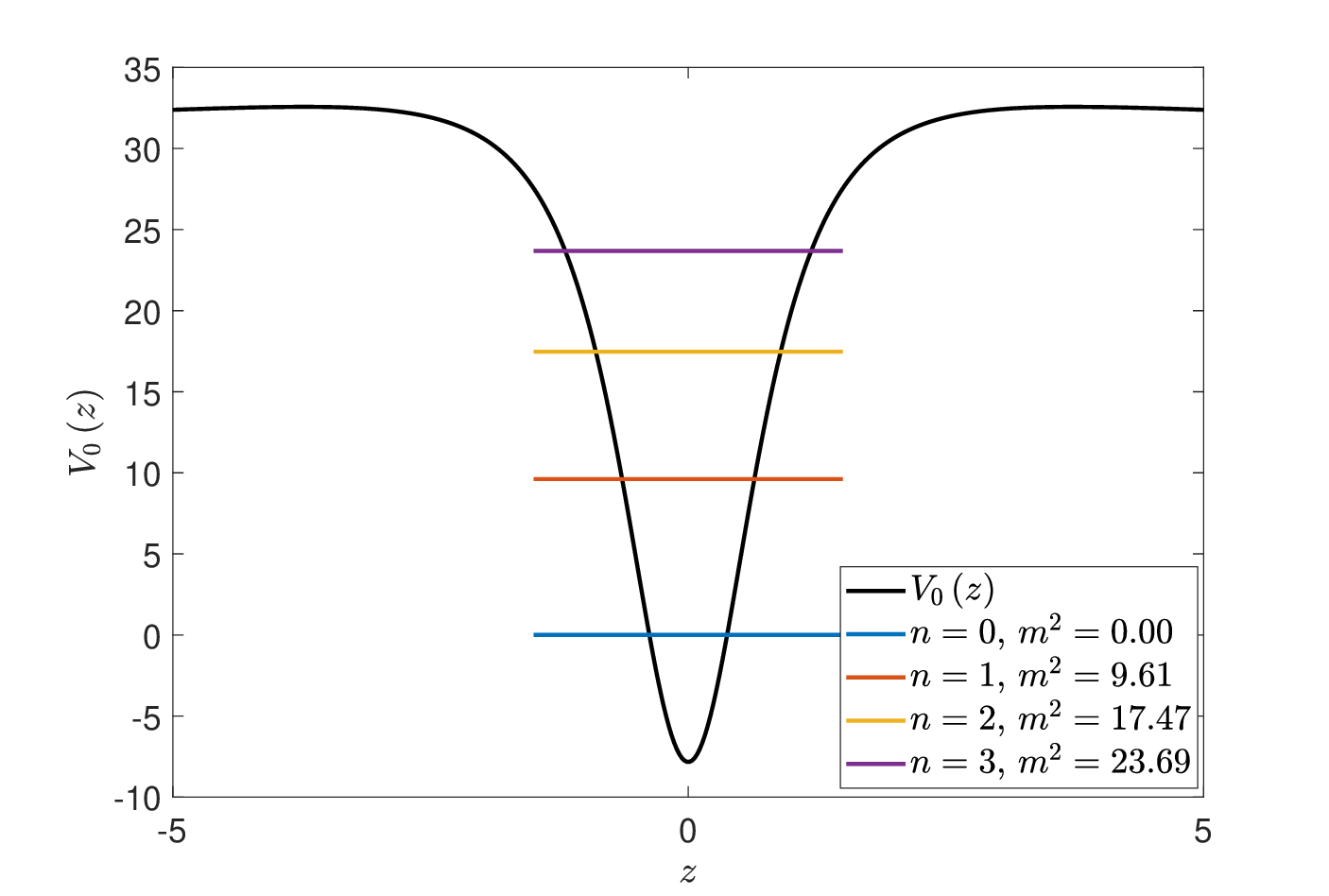}
}
\subfigure[$\text C_1=10,k=1,\text C_2=0.5$]{\label{fig:subfig2b}
\includegraphics[width=0.31\textwidth]{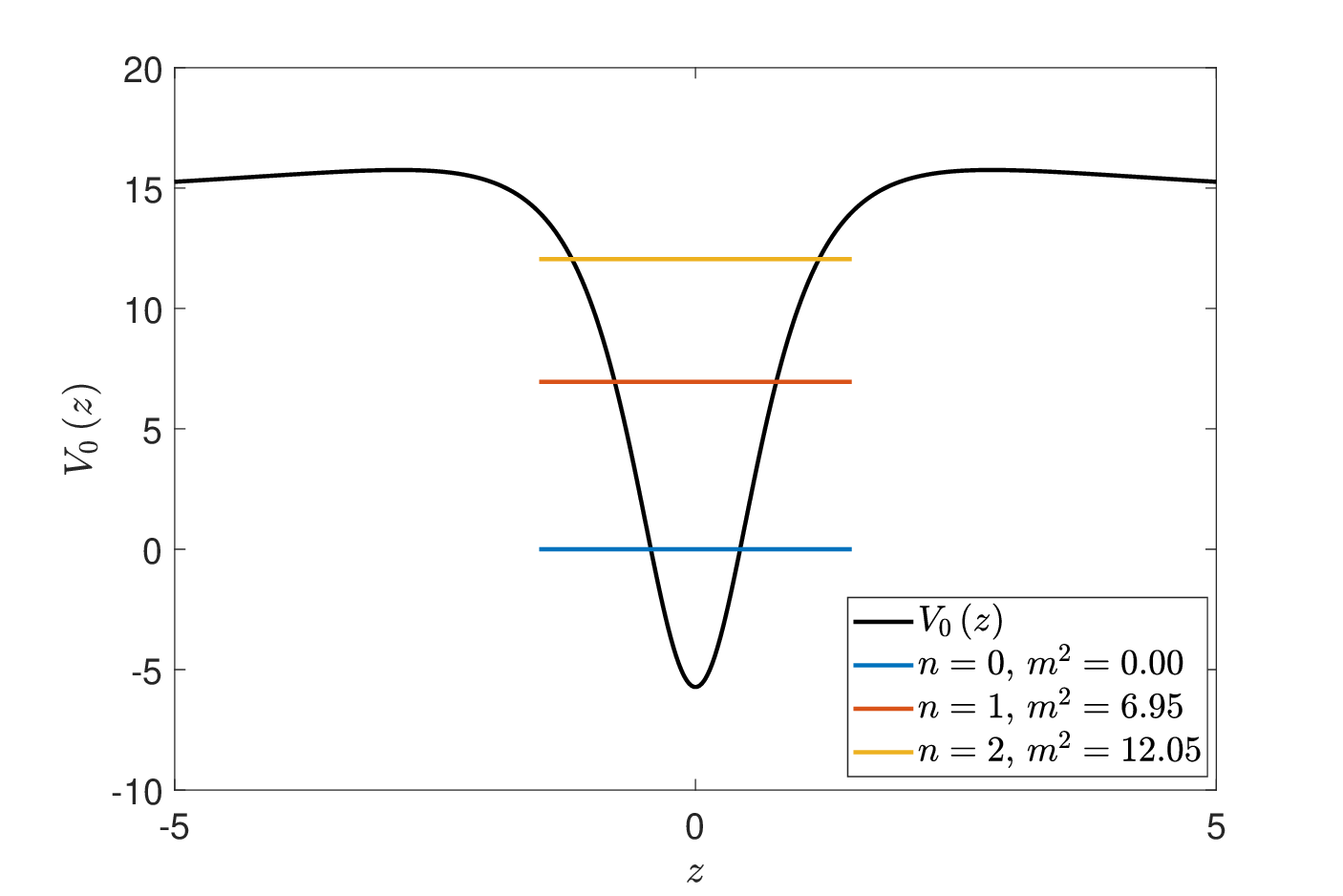}
}
\subfigure[$\text C_1=10,k=1,\text C_2=0.6$]{\label{fig:subfig2c}
\includegraphics[width=0.31\textwidth]{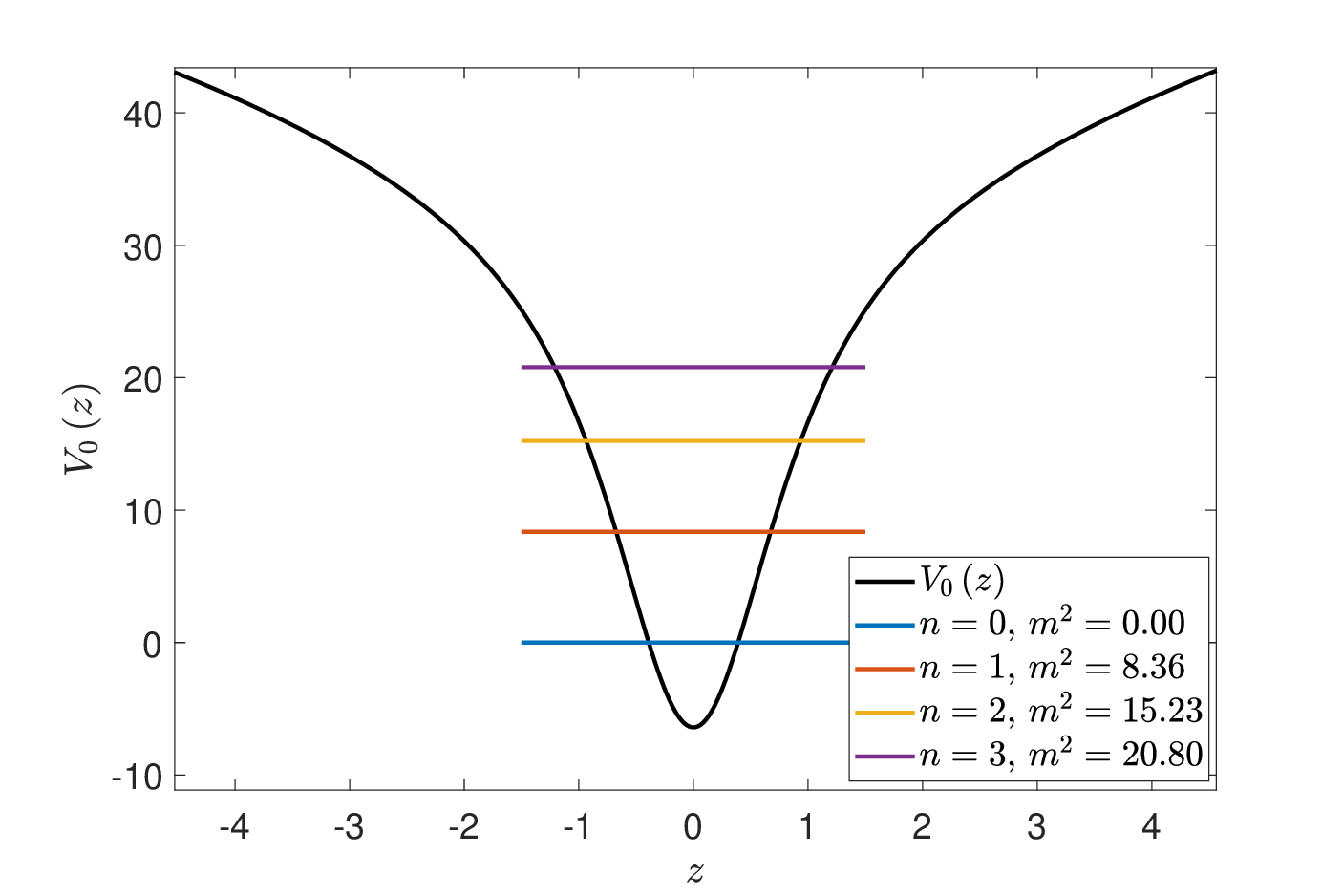}
}
\subfigure[$\text C_1=15,k=1,\text C_2=0.5$]{\label{fig:subfig2d}
\includegraphics[width=0.31\textwidth]{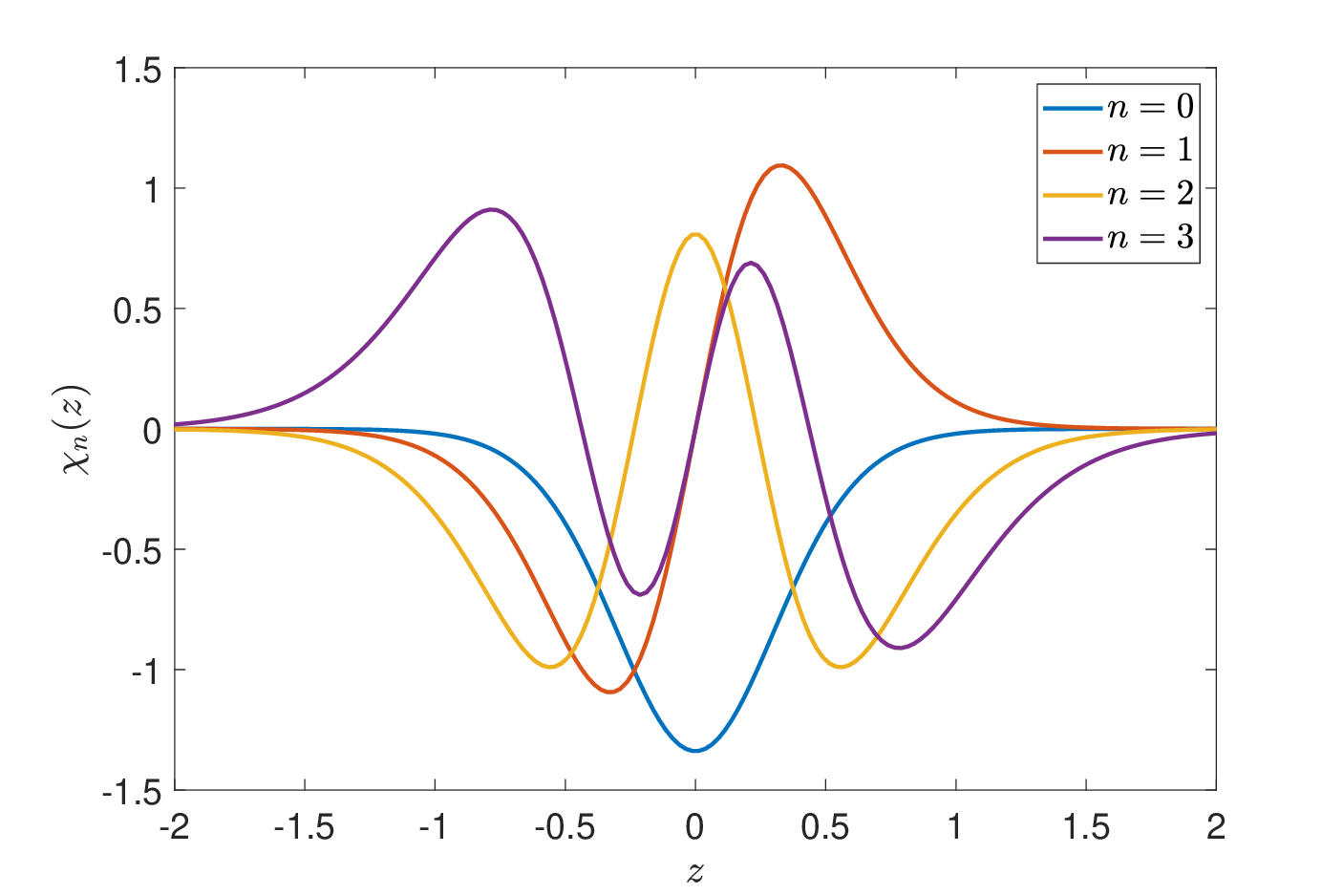}
}
\hspace{0.05cm}
\subfigure[$\text C_1=10,k=1,\text C_2=0.5$]{\label{fig:subfig2e}
\includegraphics[width=0.31\textwidth]{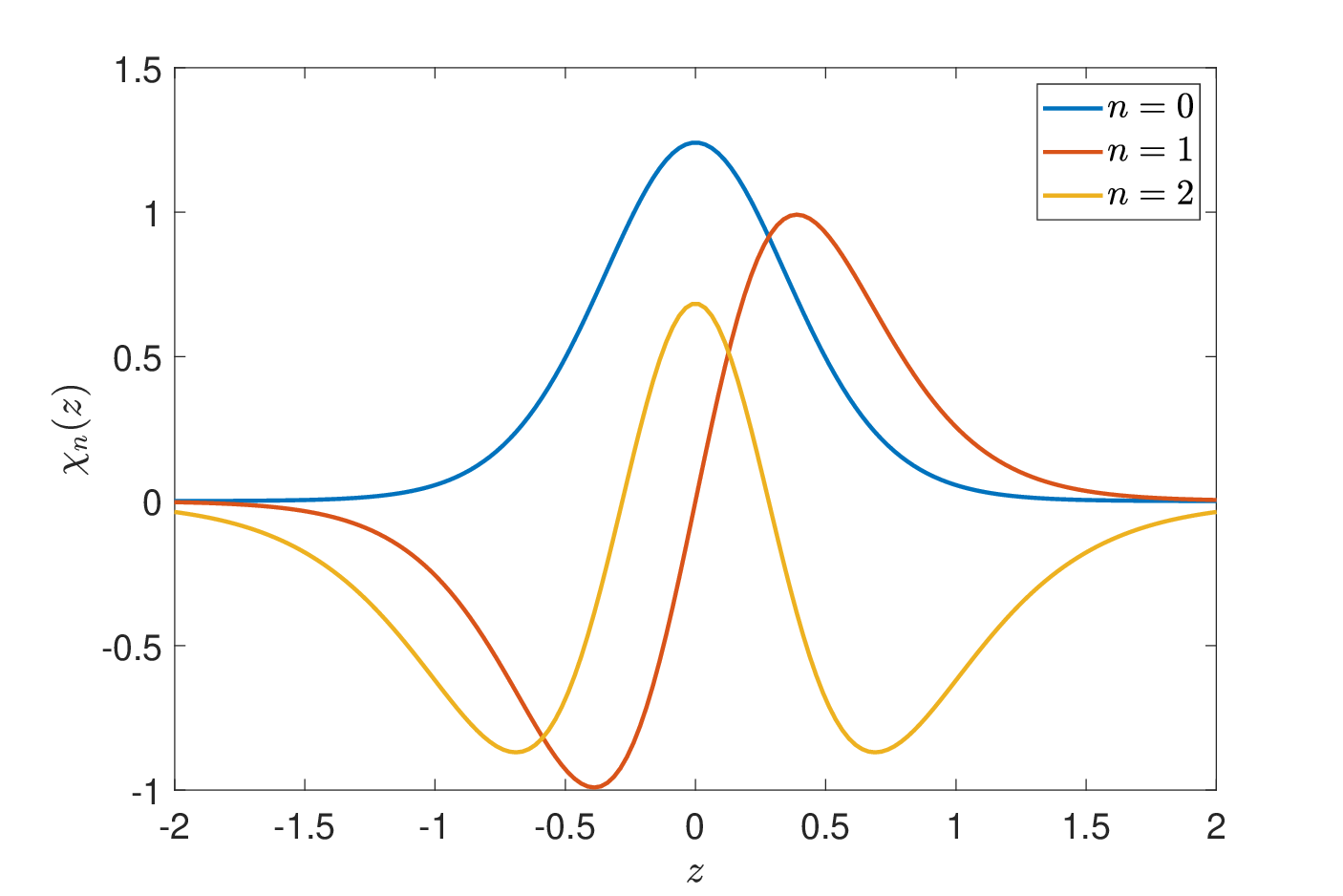}
}
\hspace{0.05cm}
\subfigure[$\text C_1=10,k=1,\text C_2=0.6$]{\label{fig:subfig2f}
\includegraphics[width=0.31\textwidth]{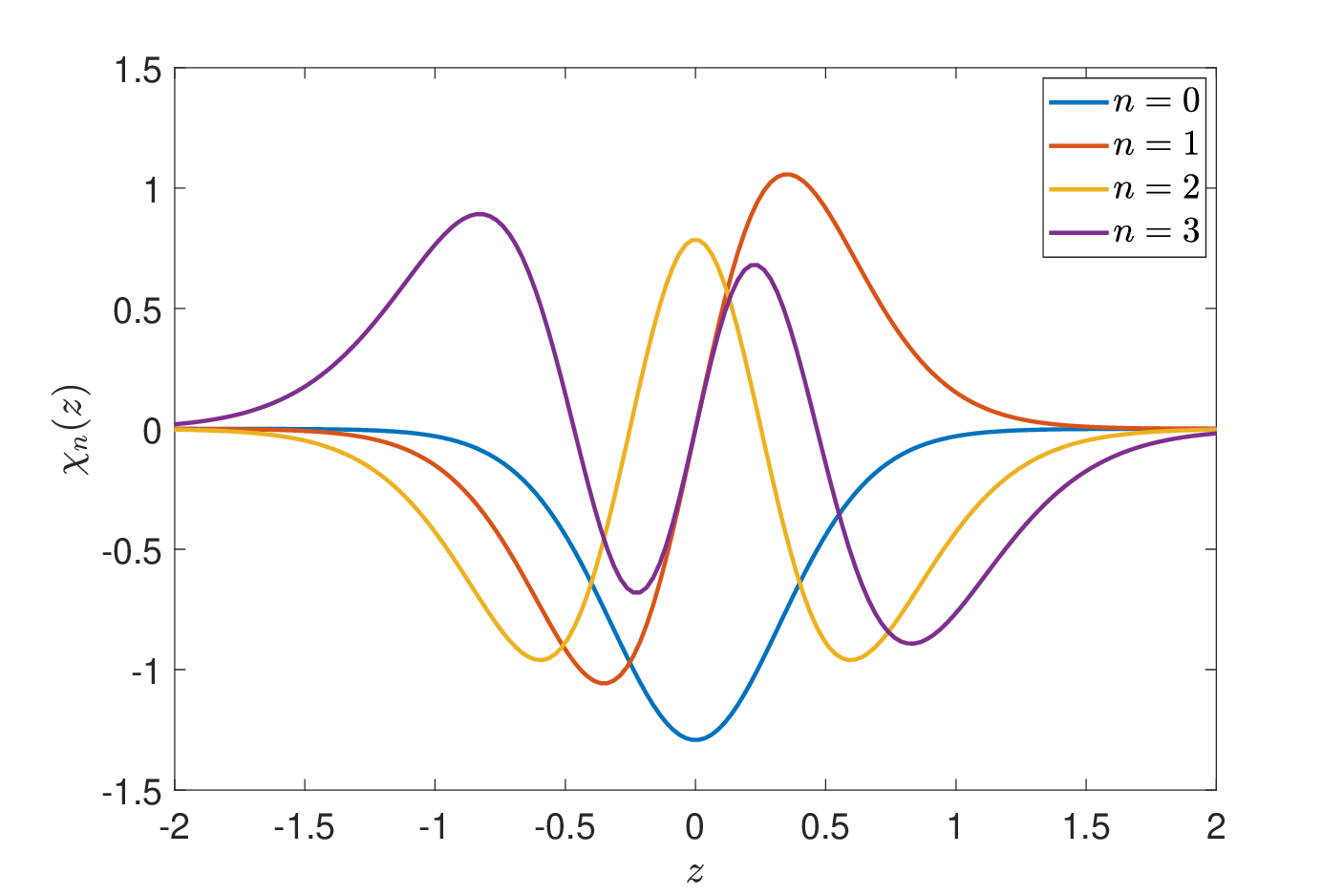}
}
\caption{The upper row shows the potential $\displaystyle V_0\left(z\right)$ with the black line representing the potential and the colored lines indicating the position of mass spectra. The lower row presents the corresponding solutions of $\displaystyle \chi_n\left(z\right)$. Parameters are varied as $\text C_1=15,10,10; k=1$ and $\text C_2=0.5,0.5,0.6$, respectively.}
\label{fig:subfig_2}
\end{figure*}

Analysis of the mass spectra for the scalar KK modes (in Fig. \ref{fig:subfig_2}) reveals that the ground state
corresponds to the zero mode (a bound state), and all the massive KK modes are bound states. Notably, the number
of bound states increases with both the parameters $\text C_1$ and $\text C_2$.

\begin{figure}[htbp]
\subfigure[$V_\varphi(z)$]{\label{fig:subfig3a}
\includegraphics[width=0.95\columnwidth]{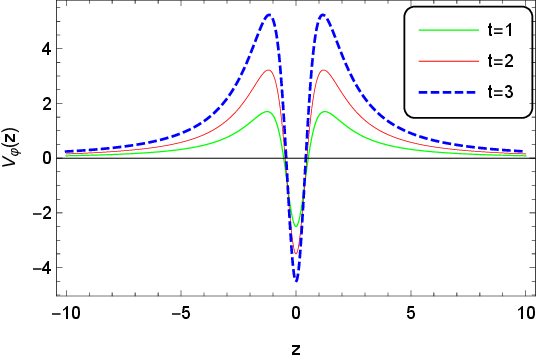}}
\subfigure[$\chi_\varphi(z)$]{\label{fig:subfig3b}
\includegraphics[width=0.95\columnwidth]{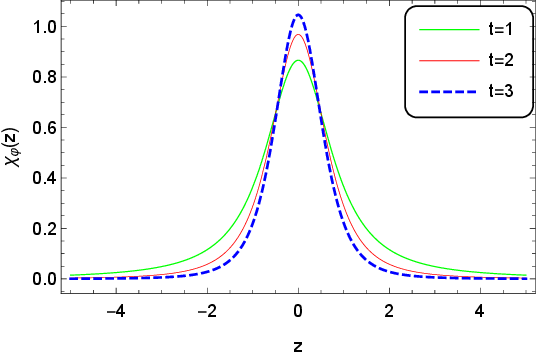}}
\caption{The effective potentials $V_\varphi(z)$ in (a), and the shapes of the scalar zero-mode
         $\chi_\varphi(z)$ in (b). The parameters are set as $\lambda=10, k=1$ and $t=1, 2, 3$.}
\label{fig:subfig_3}
\end{figure}

$Case$ II:
Coupling with the background scalar field, $\displaystyle F\left(R,\varphi\right)=\tilde F(\varphi)$. By
substituting the warp factor and the coupling function into the expressions for the effective potential (\ref{eq37})
and the zero-mode (\ref{eq41}), we obtain the zero-mode function:
\begin{equation} \label{eq49}
\chi _{\varphi}(z) = N_0\left(\frac{147}{29}\right)^{\frac{t}{2}}\lambda^{\frac12}(1+k^2z^2)^{-\frac12t-\frac34},
\end{equation}
and the potential function:
\begin{equation} \label{eq50}
V_{\varphi}\left( z \right) =\frac{k^2(3+2t)(-2+k^2(5+2t)z^2)}{4(1+k^2z^2)^2}.
\end{equation}
It is evident that
\begin{eqnarray}
\chi _{\varphi}\left( z\rightarrow\infty \right) &\propto& \frac{1}{z^{t+\frac{3}{2}}},      \label{eq51}   \\
    V_{\varphi}\left( z\rightarrow\infty \right) &\rightarrow& 0.                                \label{eq51-1}
\end{eqnarray}
Thus, the scalar zero mode is convergent, and can always be localized on the brane with scalar field coupling.
By selecting different values of parameter $t$, various potential functions and zero-mode cases can be obtained,
as depicted in Fig. \ref{fig:subfig_3}. We can see that the effective potential is a series of volcanic potentials,
and the zero-mode is still localizable.

In conclusion, the analysis of the coupling with the gravity and and the background scalar field has been
carried out separately. In the case of coupling with the gravity, it has been observed that the scalar
zero-mode can always be localized on the brane. For the effective potential, there is a critical value
$1/2$ for the parameter $\text C_2$. If $\text C_2<1/2$, the effective potential is a series of volcanic
potentials. If $\text C_2=1/2$, a finitely deep potential well emerges with the depth increasing alongside
parameters $k$ and $\text C_1$, and if $\text C_2>1/2$, there will be an infinitely deep potential well.
In the case of coupling with the background scalar field, the parameter $t$ does not affect the convergence
of the scalar zero-mode but alters the height of the potential function, with larger values of $t$ leading
to a higher potential peak.

\subsection{$\displaystyle U(1)$ gauge vector field}
In general, $\displaystyle A_{\mu}$ denotes the $\displaystyle U(1)$ gauge vector fields with $\displaystyle q=1$.
This field is utilized to describe particles with spin-1. Given that there are canonical degrees of freedom,
we initially choose the gauge $\displaystyle A_z = 0$ and subsequently perform the KK decomposition.

Considering the coupling function, the action of a 5D vector field is expressed as:
\begin{equation} \label{eq52}
S_\text{vector}=-\frac{1}{4}\int d^5x \sqrt{-g}F(R,\varphi) F_{MN}F^{MN},
\end{equation}
where the field strength $\displaystyle F_{MN}=\partial _M A_N - \partial _N A_M$. Based on the flat metric
(\ref{eq14}), the equations of motion can be decomposed into the following form:
\begin{equation} \label{eq53}
\begin{aligned}
\partial _{\mu}\left( \sqrt{-g}F(R,\varphi) F^{\mu \nu} \right) + \partial _z\left( \sqrt{-g}F(R,\varphi) F^{z\nu} \right) &= 0, \\
\partial _{\mu}\left( \sqrt{-g}F(R,\varphi) F^{\mu z} \right) &= 0.
\end{aligned}
\end{equation}
By adopting the KK decomposition
\begin{equation} \label{eq54}
A_{\mu}(x^{\mu},z) = \sum_n a_{\mu}^{(n)}(x^{\mu}) \rho_n(z) e^{-\frac{1}{2}A}F(R,\varphi)^{-\frac{1}{2}},
\end{equation}
we can obtain the following specific form of the Schr\"{o}dinger-like equation for the vector field:
\begin{equation} \label{eq55}
\left[ -\partial _{z}^{2} + V_1(z) \right] \rho_n(z) = m_{n}^{2} \rho_n(z),
\end{equation}
where the effective potential is
\begin{eqnarray} \label{eq56}
V_1(z) &=&\frac{1}{2}A''(z) + \frac{1}{4}A'^2(z) + \frac{F''(R,\varphi)}{2F(R,\varphi)}      \nonumber    \\
     & & +\frac{F'(R,\varphi)A'(z)}{2F(R,\varphi)} - \frac{F'^2(R,\varphi)}{4F^2(R,\varphi)}.
\end{eqnarray}

Similar to the case of the scalar field, the orthogonal normalization is achieved with
\begin{equation} \label{eq57}
\int \rho_m(z) \rho_n(z) dz = \delta_{mn}.
\end{equation}
Then, the 4D effective action of the vector field can be derived as:
\begin{equation} \label{eq58}
S_\text{vector} = -\frac{1}{4} \sum_n \int d^4x \sqrt{-\eta} \left( f_{\mu \nu}f^{\mu \nu} + m_{n}^{2}a_{\mu}a^{\mu} \right),
\end{equation}
where $\displaystyle f_{\mu \nu}=\partial_\mu a_\nu-\partial_\nu a_\mu$ is the 4D field strength.

By setting $\displaystyle m_n=0$, we can obtain the specific form of the vector zero mode as:
\begin{equation} \label{eq59}
\rho_{0}(z) = N_1 e^{\frac{1}{2}A(z)} F(R,\varphi)^{\frac{1}{2}},
\end{equation}
where $N_1$ is the normalization constant.

$Case$ I:
Coupling with the gravity, $\displaystyle F\left(R,\varphi\right) = G\left(R\right)$. By substituting
Eqs. (\ref{eq29}) and (\ref{eq30}) into the expressions for the potential function (\ref{eq56}) and the
zero-mode (\ref{eq59}), firstly we can obtain the vector zero-mode solution as:
\begin{equation} \label{eq60}
\rho_{0}\left( z \right) =N_1\frac{\sqrt{e^{\text C_1-(\frac{5}{7})^{\text C_2}\!\:\text C_1\!\:(\frac{1}{1+k^2\!\:z^2})^{-\text C_2}}}}
                          {(1+k^2\!\:z^2)^{\frac{1}{4}}}.
\end{equation}
If $z\rightarrow \infty$, there is
\begin{equation} \label{eq60-1}
\rho_{0}( z\rightarrow\infty) \rightarrow N_1k^{-\frac12}e^{\frac12\text C_1}z^{-\frac12}
     e^{-\frac12(\frac57)^{\text C_2}\text C_1k^{2\text C_2}z^{2\text C_2}}.
\end{equation}
As parameters $\text C_1,\text C_2$ are positive, this vector zero-mode converges to zero exponentially
when far away from the brane. Therefore, the zero-mode is still normalizable, and can always be localized
on the thick brane.

Furthermore, the effective potential (\ref{eq62}) can be expressed as :
\begin{eqnarray} \label{eq62}
V_1\left( z \right) \hspace{-0.1cm}&=&\hspace{-0.1cm} \frac{1}{4}\!\:k^2\!\:\bigg(\frac{1}{1+k^2\!\:z^2}\bigg)^{2-2\!\:\text C_2}\!\:
      \bigg[4\!\:(\frac{25}{49})^{\text C_2}\!\:{\text C_1}^2\!\:{\text C_2}^2\!\:k^2\!\:z^2                                  \nonumber  \\
   & &\hspace{-0.1cm} +\bigg(\frac{1}{1+k^2\!\:z^2}\bigg)^{2\!\:\text C_2}
       \!\:(-2+3\!\:k^2\!\:z^2)-4\times \!\:5^{\text C_2}\text C_1\text C_2         \nonumber  \\
   & &\hspace{-0.1cm} \times\!\:\bigg(\frac{1}{7+7\!\:k^2\!\:z^2}\bigg)^{\text C_2}\!\:(1+2\!\:(-1+\text C_2)\!\:k^2\!\:z^2)\bigg].
\end{eqnarray}
For an analysis of the specific behavior of the potential function at infinity, we approximate:
\begin{equation} \label{eq64}
V_1( z\rightarrow \infty )\rightarrow \left(\frac{25}{49}\right)^{\text C_2}\text C_1^2\text C_2^2k^{4\text C_2}z^{4\text C_2-2}.
\end{equation}
From the above expression, we
can derive the conditions for the value of $\displaystyle V_1\left( z \right) $ on the boundary, namely,
\begin{equation} \label{eq65}
V_1\left( z\rightarrow\pm \infty \right) \rightarrow\begin{cases}
	+\infty,&		\text C_2>1/2\\
	V_{\text{const}},&		\text C_2=1/2\\
	0,&		0<\text C_2<1/2,\\
\end{cases}
\end{equation}
where $\displaystyle V_{\text{const}}=\frac{5}{28}\text C_1^2k^2$.

\begin{figure}[htbp]
\subfigure[$V_1\left(z\right)$]{\label{fig:subfig4a}
\includegraphics[width=0.95\columnwidth]{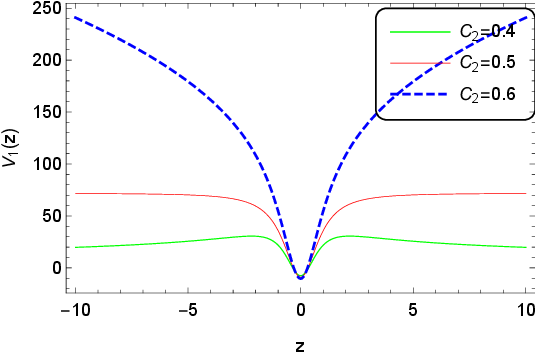}}
\subfigure[$\rho_0\left(z\right)$]{\label{fig:subfig4b}
\includegraphics[width=0.95\columnwidth]{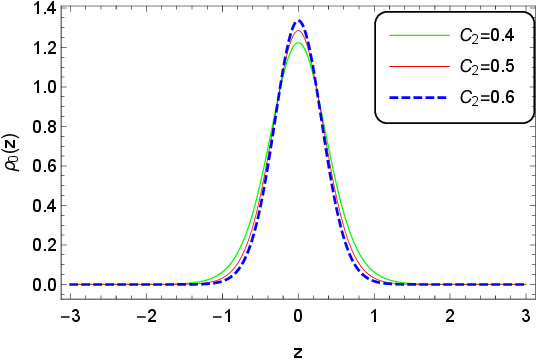}}
\caption{The effective potentials $\displaystyle V_1\left(z\right)$ in (a), and the shapes
         of the vector zero-mode $\displaystyle \rho_0\left(z\right)$ in (b). The parameters
         are set as $\text C_1=20, k=1$, and $\text C_2=0.4, 0.5, 0.6$.}
\label{fig:subfig_4}
\end{figure}

\begin{figure*}[htbp]
\subfigure[$\text C_1=15,k=1,\text C_2=0.5$]{\label{fig:subfig5a}
\includegraphics[width=0.31\textwidth]{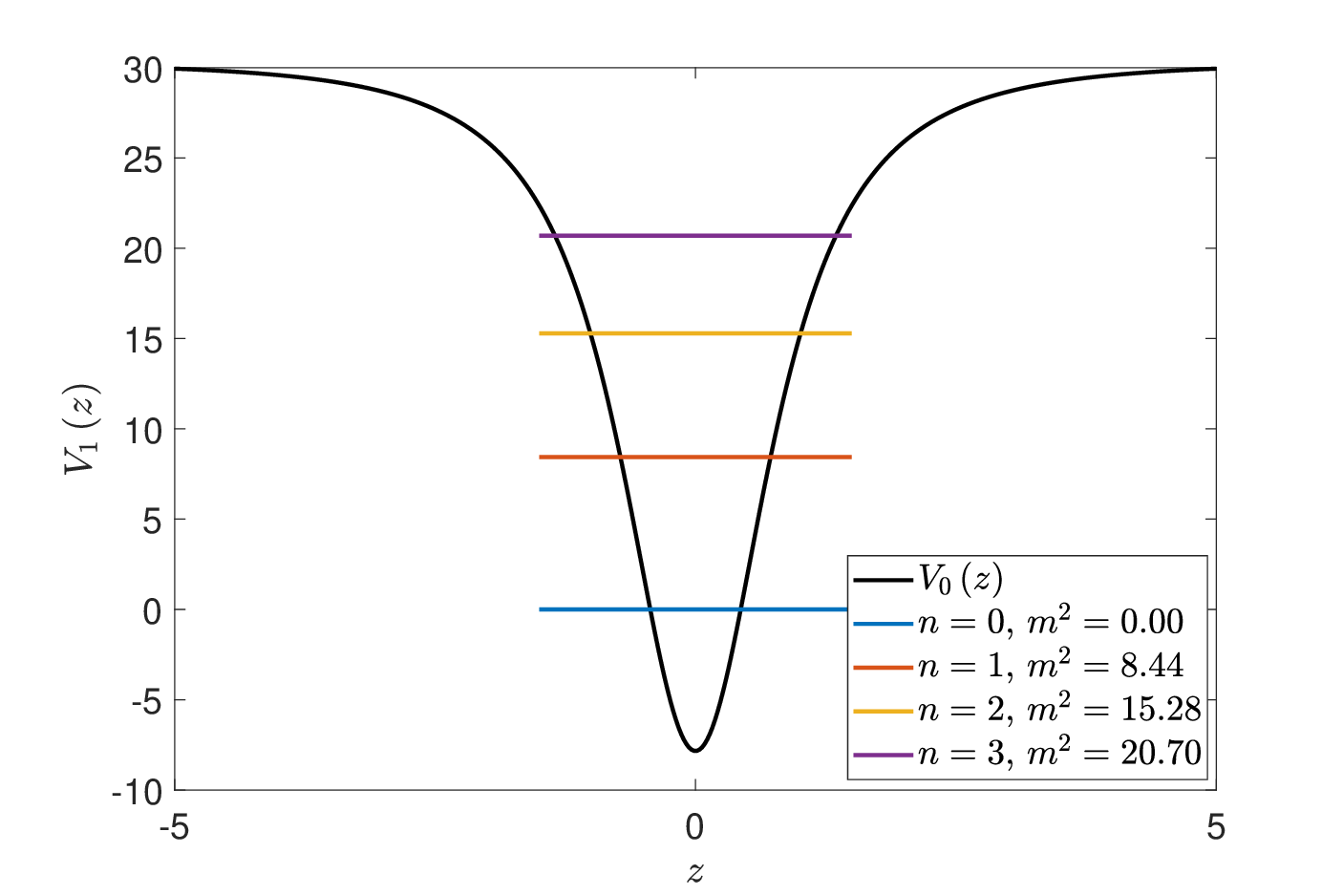}}
\subfigure[$\text C_1=10,k=1,\text C_2=0.5$]{\label{fig:subfig5b}
\includegraphics[width=0.31\textwidth]{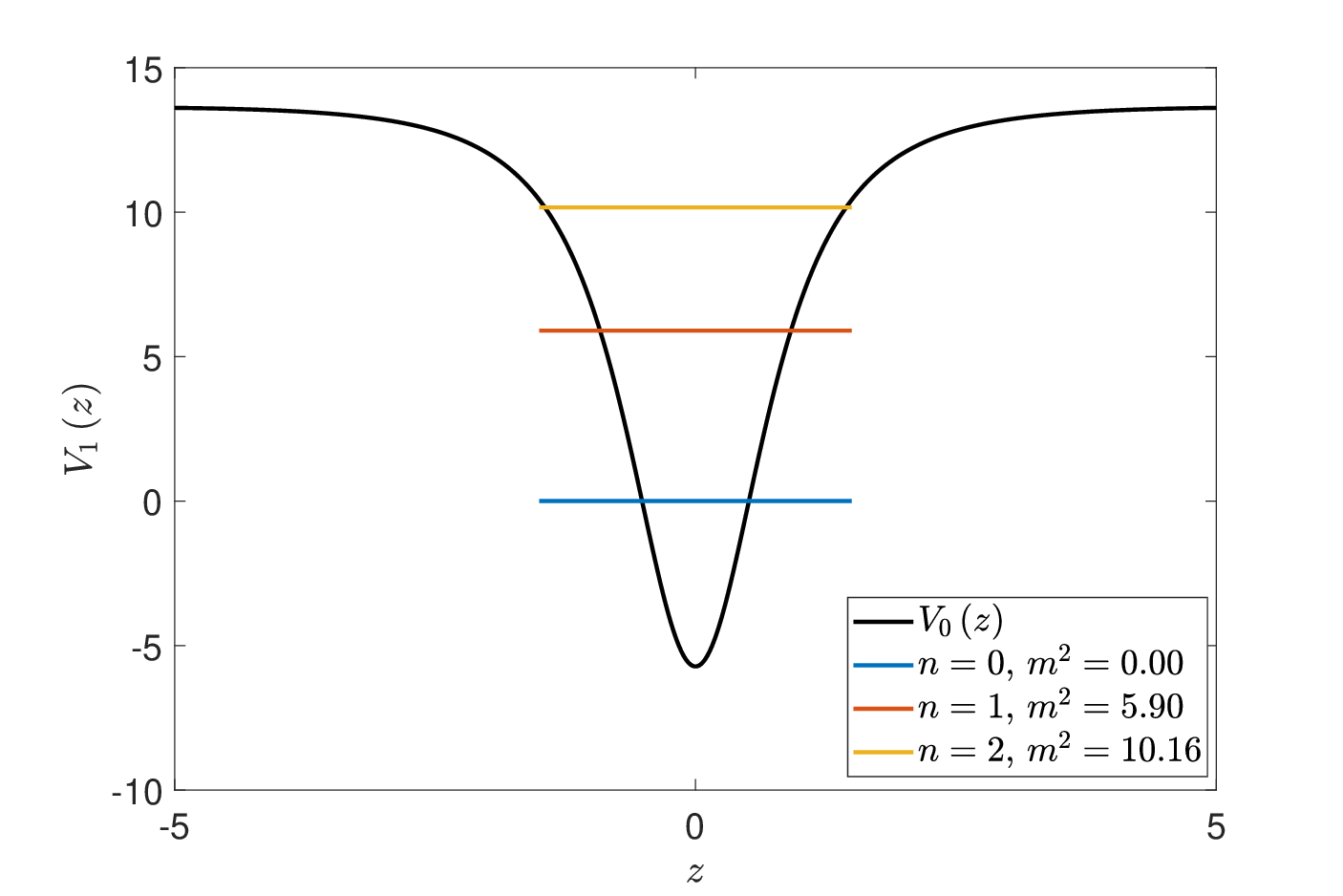}}
\subfigure[$\text C_1=10,k=1,\text C_2=0.6$]{\label{fig:subfig5c}
\includegraphics[width=0.31\textwidth]{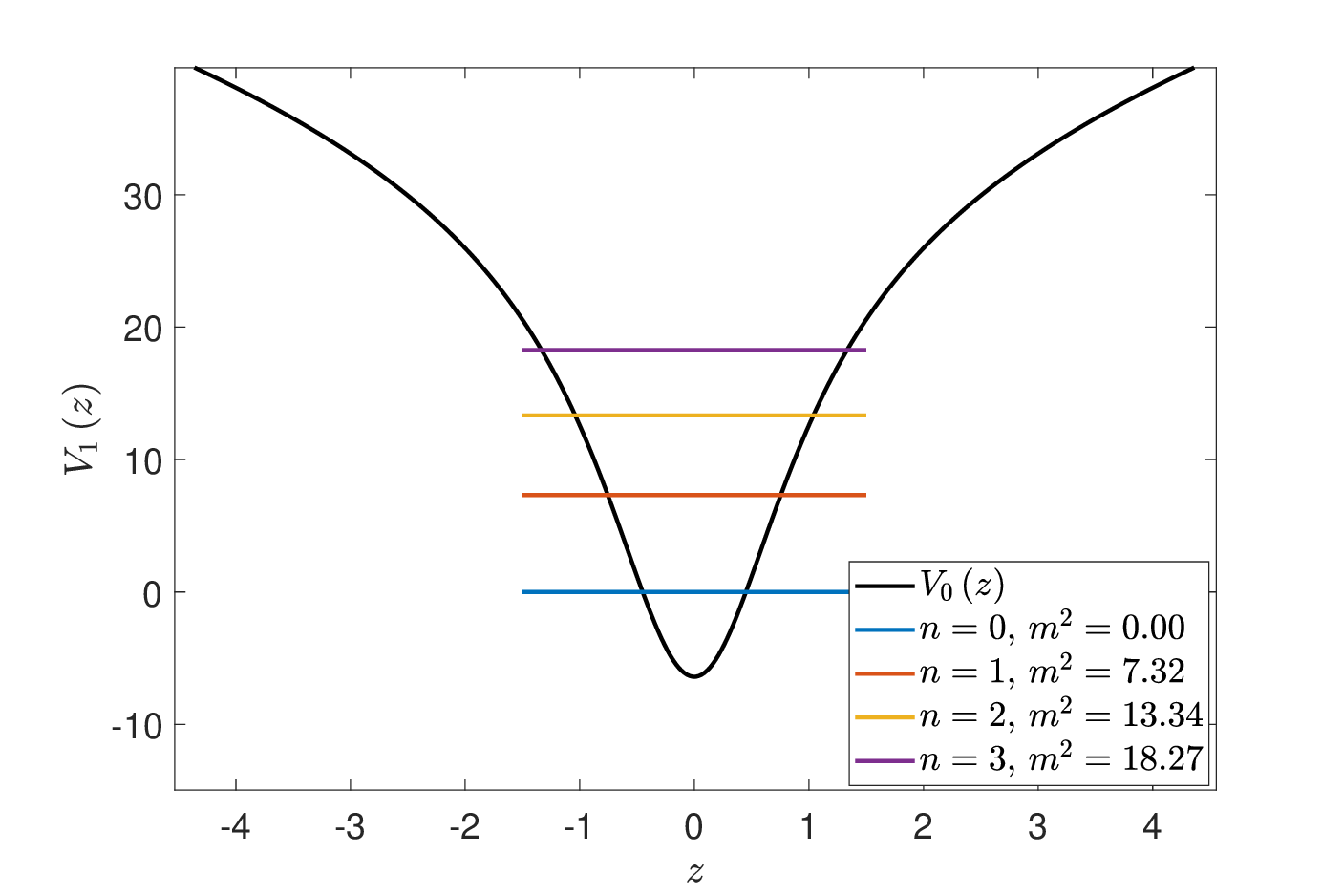}}
\subfigure[$\text C_1=15,k=1,\text C_2=0.5$]{\label{fig:subfig5d}
\includegraphics[width=0.31\textwidth]{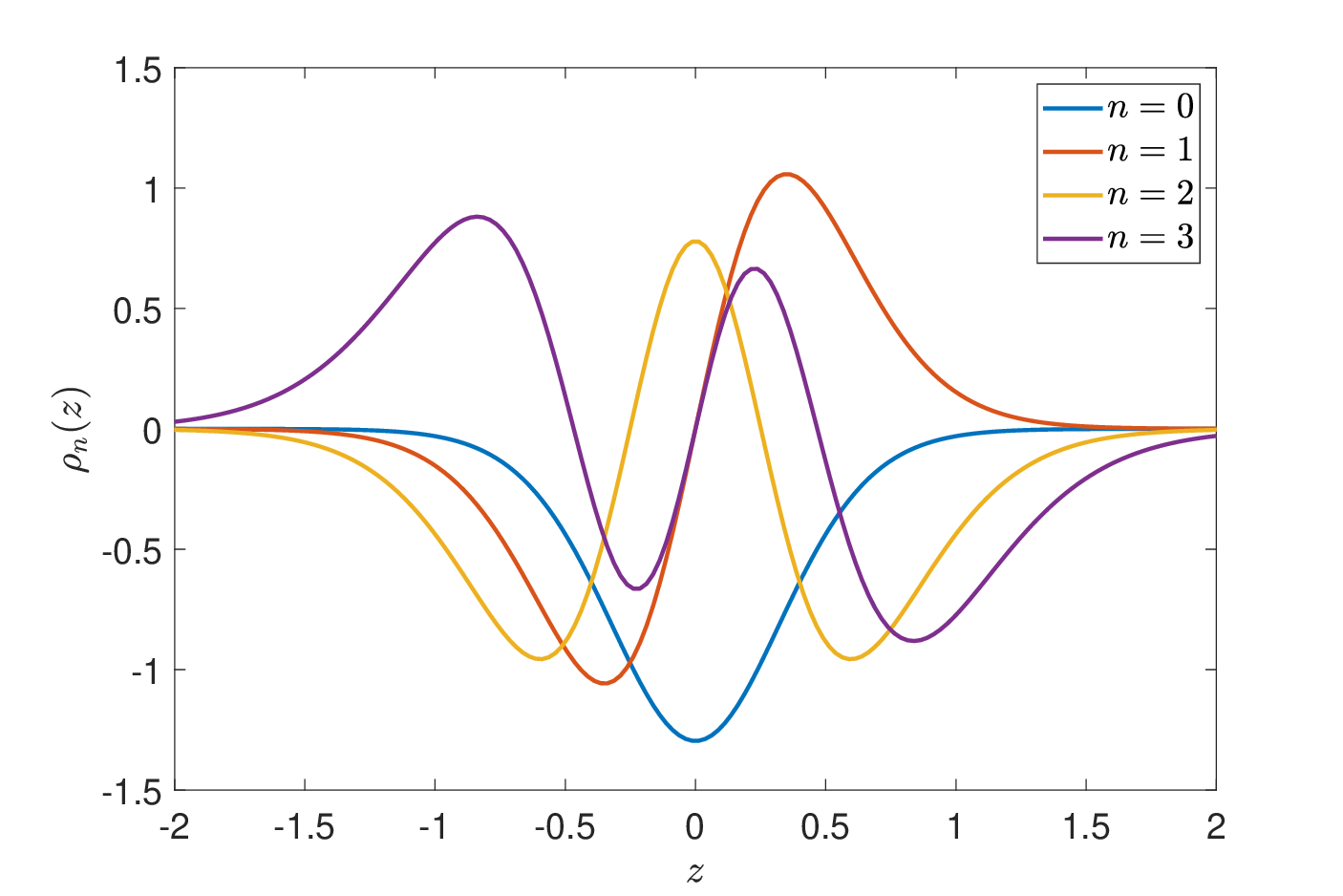}}
\hspace{0.1cm}
\subfigure[$\text C_1=10,k=1,\text C_2=0.5$]{\label{fig:subfig5e}
\includegraphics[width=0.31\textwidth]{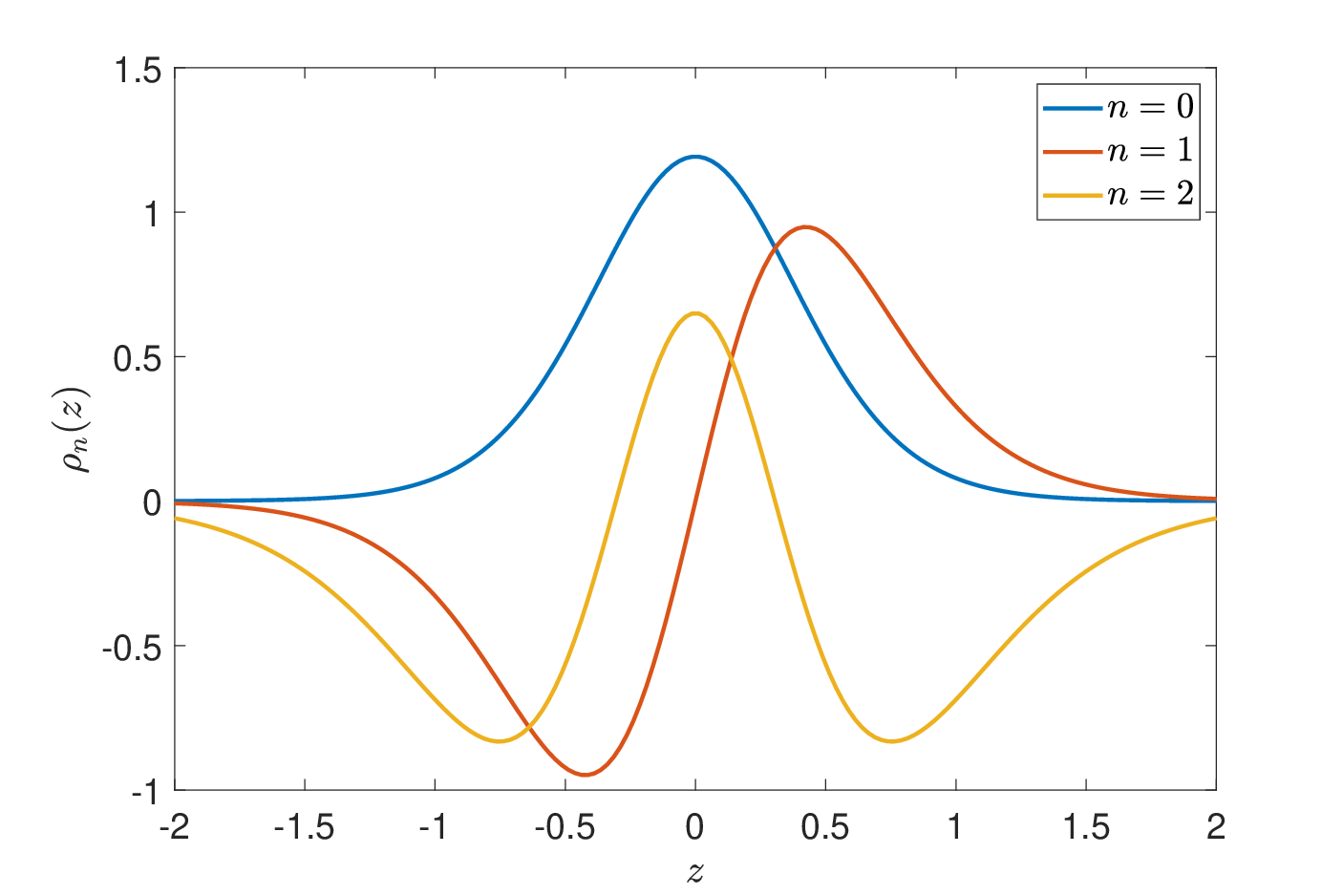}}
\hspace{0.1cm}
\subfigure[$\text C_1=10,k=1,\text C_2=0.6$]{\label{fig:subfig5f}
\includegraphics[width=0.31\textwidth]{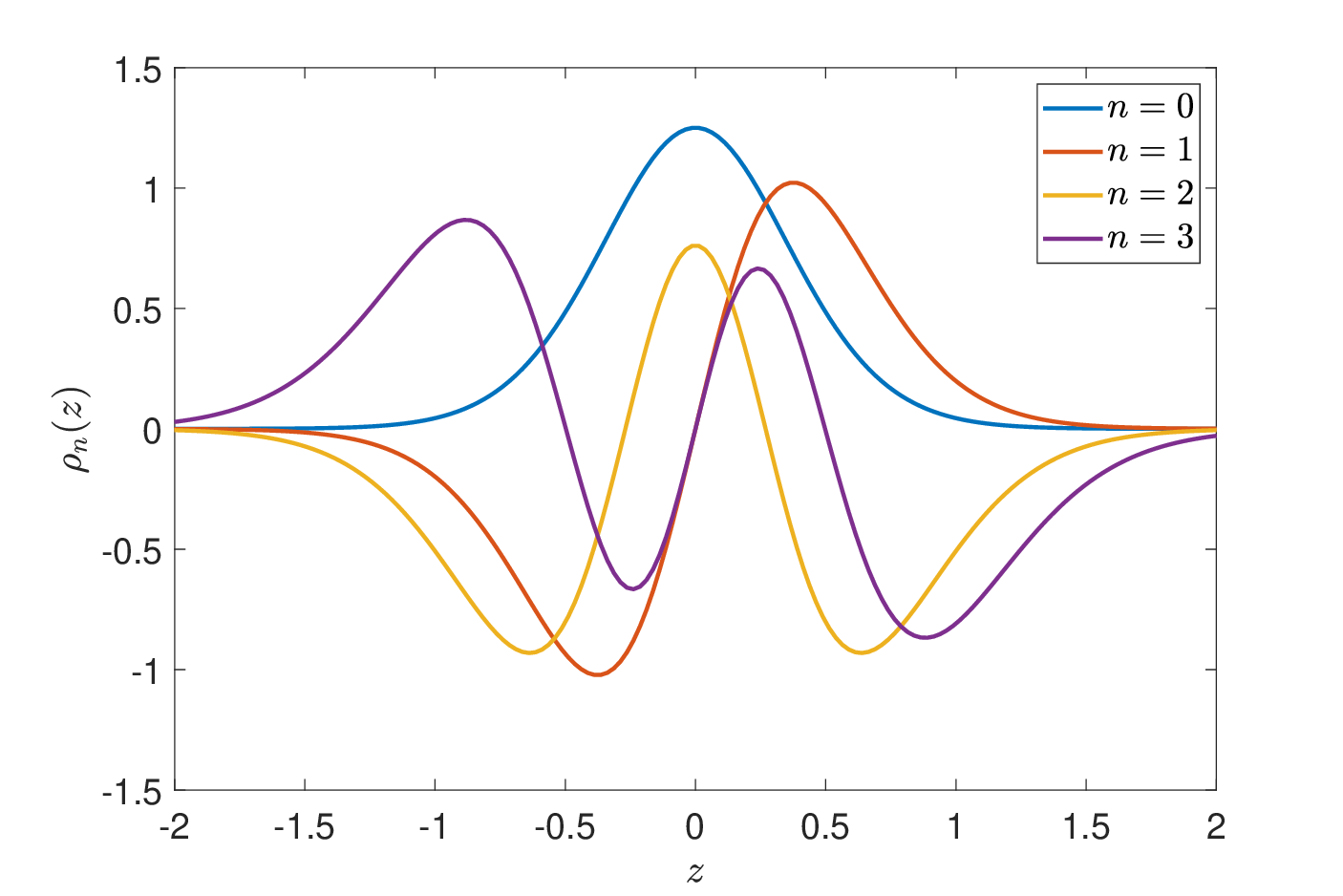}}
\caption{The upper row shows the potential $\displaystyle V_1\left(z\right)$ with the black line
         representing the potential and the colored lines indicating the position of mass spectra.
         The lower row presents the corresponding solutions of $\displaystyle \rho_n\left(z\right)$.
         Parameters are varied as $\text C_1=15,10,10; k=1$ and $\text C_2=0.5,0.5,0.6$, respectively.}
\label{fig:subfig_5}
\end{figure*}

To more intuitively reflect the above analysis, we plot the effective potential $V_1(z)$ and the zero-mode
$\rho_0(z)$ in Fig. \ref{fig:subfig_4}. The zero mode can always be localized on the brane. Furthermore,
the condition $\text C_2=1/2$ means a finite number of localized massive modes, if $\text C_2>1/2$, there will be
infinite number of localized massive modes, and if $0<\text C_2<1/2$, no massive mode can be localized on the
brane. For the first two conditions, setting different values of parameters $\text C_1$ and $\text C_2$, we
solve for the bound states and illustrate them in Fig. \ref{fig:subfig_5}, alongside with the corresponding
effective potential.

$Case$ II:
Coupling with the background scalar field, $\displaystyle F\left(R,\varphi\right)=\tilde F(\varphi)$.
Substituting equations (\ref{eq29}) and (\ref{eq30}) into the expressions for the potential function
(\ref{eq56}) and the zero-mode (\ref{eq59}) yields the zero-mode function:
\begin{equation} \label{eq67}
\rho _{\varphi}(z)= N_1\left(\frac{147}{29}\right)^{\frac12t}\lambda^{\frac12}(1+k^2z^2)^{-\frac12t-\frac14},
\end{equation}
and the effective potential:
\begin{equation} \label{eq68}
V_{\varphi}\left( z \right) =\frac{k^2(1+2t)(-2+k^2(3+2t)z^2)}{4(1+k^2z^2)^2}.
\end{equation}
It is evident that:
\begin{eqnarray}
\rho _{\varphi}\left( z\rightarrow\infty \right) &\propto& \frac{1}{z^{t+\frac{1}{2}}},    \label{eq69}    \\
 V_{\varphi}\left( z\rightarrow\infty \right) &\rightarrow& 0.      \label{eq69-1}
\end{eqnarray}
Therefore, if $\displaystyle t>1/2$, the vector zero-mode converges to zero faster than $1/z$ when
$z\rightarrow\infty$, and can be localized on the brane.

\begin{figure}[htbp]
\subfigure[$V_\varphi\left(z\right)$]{\label{fig:subfig6a}
\includegraphics[width=0.95\columnwidth]{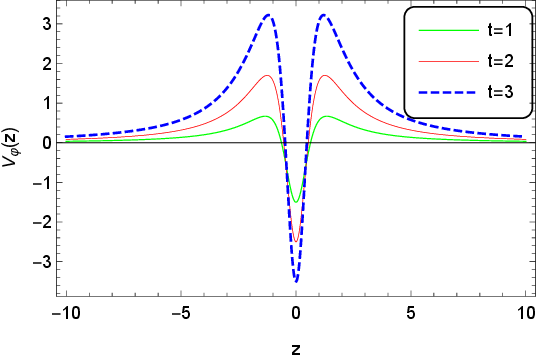}}
\subfigure[$\chi_\varphi\left(z\right)$]{\label{fig:subfig6b}
\includegraphics[width=0.95\columnwidth]{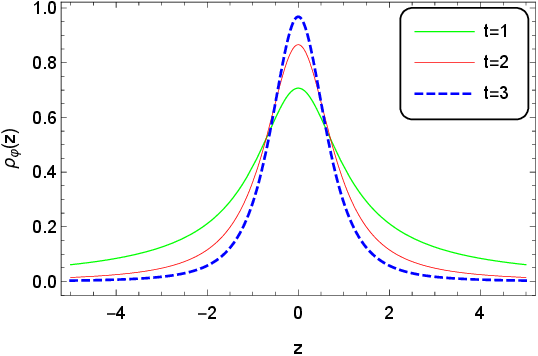}}
\caption{The effective potentials $\displaystyle V_\varphi\left(z\right)$ in (a), and the shapes of the vector zero-mode $\displaystyle \rho_\varphi\left(z\right)$ in (b). The parameters are set as $\displaystyle \lambda=10, k=1$ and $\displaystyle t=1, 2, 3$.}
\label{fig:subfig_6}
\end{figure}

By selecting different values of $t$, the effective potential and the vector zero-mode can be obtained,
as depicted in Fig. \ref{fig:subfig_6}. It can be seen that as $t>1/2$ ,the vector zero-mode is always
localized on the brane. Besides, the effective potential exhibits a higher barrier when with a larger
value of parameter $t$.

In summary, the analysis of coupling with the gravity and the background scalar field reveals distinct
influences on the localization of the zero-mode, and the behaviors of the effective potential at infinity.
In the gravity coupling case, the vector zero-mode can always be localized on the brane, and the different
values of parameter $\text C_2$ give rise to varied behaviors of the effective potential at infinity. Besides,
when the effective potential approaches a finite value ($\text C_2=1/2$), both the parameters $\text C_1$
and $k$ will influence the depth of the potential well, where larger values lead to deeper potential wells
and an increase in the number of bound states. Then, with coupling to the background scalar field, the
effective potential is in volcano form, while the localization of the vector zero mode requires the parameter
$t>1/2$.

\subsection{Kalb-Ramond field}
In the case where $\displaystyle q = 2$, the $q$-form field manifests as a KR field. This second-order
antisymmetric tensor field, initially introduced in string theory, emerges as a set of massless
excitations of a closed string. In 4D spacetime, the KR field is dual to a scalar field. However, within
higher-dimensional spacetime, it signifies the presence of new particles. Therefore, the study of its
localization in braneworld theories becomes imperative.

With considering the coupling function, the action for a free KR field in 5D spacetime is given by:
\begin{equation} \label{eq70}
S_\text{KR}=-\int{d^5x\sqrt{-g}F\left( R,\varphi \right) H_{MNL}H^{MNL}},
\end{equation}
where $\displaystyle H_{MNL}=\partial _{[M}B_{NL]}$ denotes the field strength of the KR field. Employing
the metric (\ref{eq14}), we can obtain the equations of motion:
\begin{equation} \label{eq71}
\begin{aligned}
\partial _{\mu}\left( \sqrt{-g}F\left( R,\varphi \right) H^{\mu \nu \gamma} \right)
           +\partial _z\left( \sqrt{-g}F\left( R,\varphi \right) H^{z\nu \gamma} \right) =0, \\
\partial _{\mu}\left( \sqrt{-g}F\left( R,\varphi \right) H^{\mu \nu z} \right) =0.
\end{aligned}
\end{equation}
By selecting the gauge $\displaystyle B_{\mu5}=0$ and utilizing the KK decomposition:
\begin{equation} \label{eq72}
B_{\mu \nu}\left( x^{\mu},z \right) =\sum_n{\hat{b}_{\mu \nu}^{\left( n \right)}\left( x^{\mu} \right) u^{\left( n \right)}\left( z \right)}\text{e}^{\frac{1}{2}A}F\left( R,\varphi \right) ^{-\frac{1}{2}},
\end{equation}
we derive a Schr\"{o}dinger-like equation for the KR field:
\begin{equation} \label{eq73}
\left[ -\partial _{z}^{2}+V_2\left( z \right) \right]{u}^{\left( n \right)}\left( z \right) =m_{n}^{2}{u}^{\left( n \right)}\left( z \right),
\end{equation}
where the effective potential is
\begin{eqnarray} \label{eq74}
V_2\left( z \right) &=&  -\frac{1}{2}A''(z)+\frac{1}{4}A'^2\left( z \right)
                    +\frac{F''\left( R,\varphi \right)}{2F\left( R,\varphi \right)}                \nonumber  \\
      & &-\frac{F'\left( R,\varphi \right)}{2F\left( R,\varphi \right)}A'\left( z \right)
                    -\frac{F'^2\left( R,\varphi \right)}{4F^2\left( R,\varphi \right)}.
\end{eqnarray}
Assuming orthogonal normalization:
\begin{equation} \label{eq75}
\int{u_m\left( z \right) u_n\left( z \right) dz=\delta _{mn}},
\end{equation}
the 4D effective action, deduced from the 5D one, is:
\begin{equation} \label{eq76}
S_\text{KR}=\sum_n{\int{d^4x\sqrt{-\eta}(\hat{h}_{\mu \nu \gamma}^{\left( n \right)}\hat{h}^{\left( n \right) \mu \nu \gamma}+\frac{1}{3}m_{n}^{2}\hat{b}_{\mu \nu \gamma}^{\left( n \right)}\hat{b}^{\left( n \right) \mu \nu \gamma})}}.
\end{equation}
For $\displaystyle m_{n}^{2}=0$, the zero-mode solution of the KR field can be resolved from equation (\ref{eq73}):
\begin{equation} \label{eq77}
u_{0}\left({z} \right) =N_2\text{e}^{-\frac{1}{2}\!\:A\left( z \right)}\!\:F\left( R,\varphi\right) ^{\frac{1}{2}}
\end{equation}
with $N_2$ the normalization constant.

$Case$ I:
Coupling with the gravity, $\displaystyle F(R,\varphi) = G(R)$. By substituting Eqs. (\ref{eq29}) and (\ref{eq30})
into the effective potential (\ref{eq74}) and the zero-mode expression (\ref{eq77}), firstly we can obtain the
expression for the zero-mode
\begin{equation} \label{eq78}
u_{0}\left( z \right) =N_2\sqrt{\text{e}^{\text C_1-(\frac{5}{7})^{\text C_2}\!\:\text C_1(\frac{1}{1+k^2\!\:z^2})^{-\text C_2}}}
                       \!\:(1+k^2\!\:z^2)^{\frac14}.
\end{equation}
For this zero-mode, if $z\rightarrow\infty$, there is
\begin{equation} \label{eq79}
u_{0}(z\rightarrow\infty)\rightarrow N_2 e^{\frac12\text C_1}k^{\frac12}z^{\frac12}e^{-\frac12(\frac57)^{\text C_2}\text C_1k^{2\text C_2}z^{2\text C_2}}.
\end{equation}
As parameters $\text C_1,\text C_2$ are positive, this zero-mode will be suppressed to zero exponentially
when far away from the brane. Therefore, the zero-mode can always be localized on the brane.

Then, the effective potential (\ref{eq74}) can be given by:
\begin{eqnarray} \label{eq80}
V_2\left( z \right)\hspace{-0.1cm} &=&\hspace{-0.1cm} \frac{1}{4}k^2\bigg(\frac{1}{1+k^2z^2}\bigg)^{2-2\text C_2}
              \bigg[4\left(\frac{25}{49}\right)^{\text C_2}{\text C_1}^2{\text C_2}^2k^2z^2        \nonumber   \\
      & &\hspace{-0.1cm}-\left(\frac{1}{1+k^2z^2}\right)^{2\text C_2}(-2+k^2z^2)-4 \times 5^{\text C_2}\text C_1\text C_2       \nonumber   \\
      & &\hspace{-0.1cm}\times\left(\frac{1}{7+7k^2z^2}\right)^{\text C_2}(1+2\text C_2k^2z^2)\bigg].
\end{eqnarray}
For a boundary analysis of the specific behavior of the effective potential at infinity, we get
\begin{equation} \label{eq82}
V_2(z\rightarrow\infty)\rightarrow \left(\frac{25}{49}\right)^{\text C_2}\text C_1^2\text C_2^2k^{4\text C_2}z^{4\text C_2-2}.
\end{equation}

Therefore, we can further derive the asymptotic behaviors of the potential \(V_2\left( z \right)\) when far away
from the brane, reads
\begin{equation} \label{eq83}
V_2\left( z\rightarrow\pm \infty \right) \rightarrow
\begin{cases}
	+\infty, & \text C_2>1/2\\
	V_{\text{const}}, & \text C_2=1/2\\
	0, & 0<\text C_2<1/2,
\end{cases}
\end{equation}
where \(V_{\text{const}}=\frac{5}{28}\text C_1^2k^2\). Clearly, \(V_{\text{const}}>0\), and its value is dependent on the parameters
\(\text C_1\) and \(k\). As $\text C_2=1/2$, the larger the values of \(\text C_1\) and \(k\), the larger \(V_{\text{const}}\) becomes,
the deeper the potential well, and the more bound states can be obtained.

\begin{figure}[htbp]
\subfigure[$V_2\left(z\right)$]{\label{fig:subfig7a}
\includegraphics[width=0.95\columnwidth]{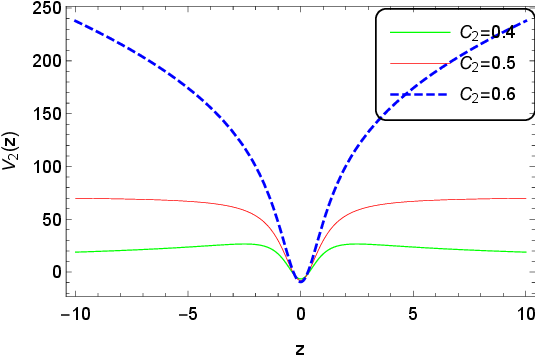}}
\subfigure[$u_0\left(z\right)$]{\label{fig:subfig7b}
\includegraphics[width=0.95\columnwidth]{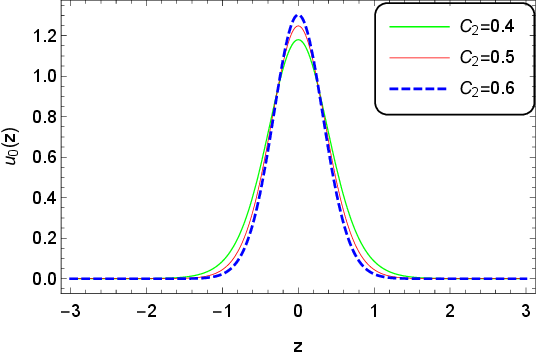}}
\caption{The effective potentials $\displaystyle V_2\left(z\right)$ in (a), and the zero-mode
         of the KR field $\displaystyle u_0\left(z\right)$ in (b). The parameters are set as
         $\text C_1=20,k=1$ and $\text C_2=0.4,0.5,0.6$.}
\label{fig:subfig_7}
\end{figure}

\begin{figure*}[htbp]
\subfigure[$\text C_1=20,k=1,\text C_2=0.5$]{\label{fig:subfig8a}
\includegraphics[width=0.31\textwidth]{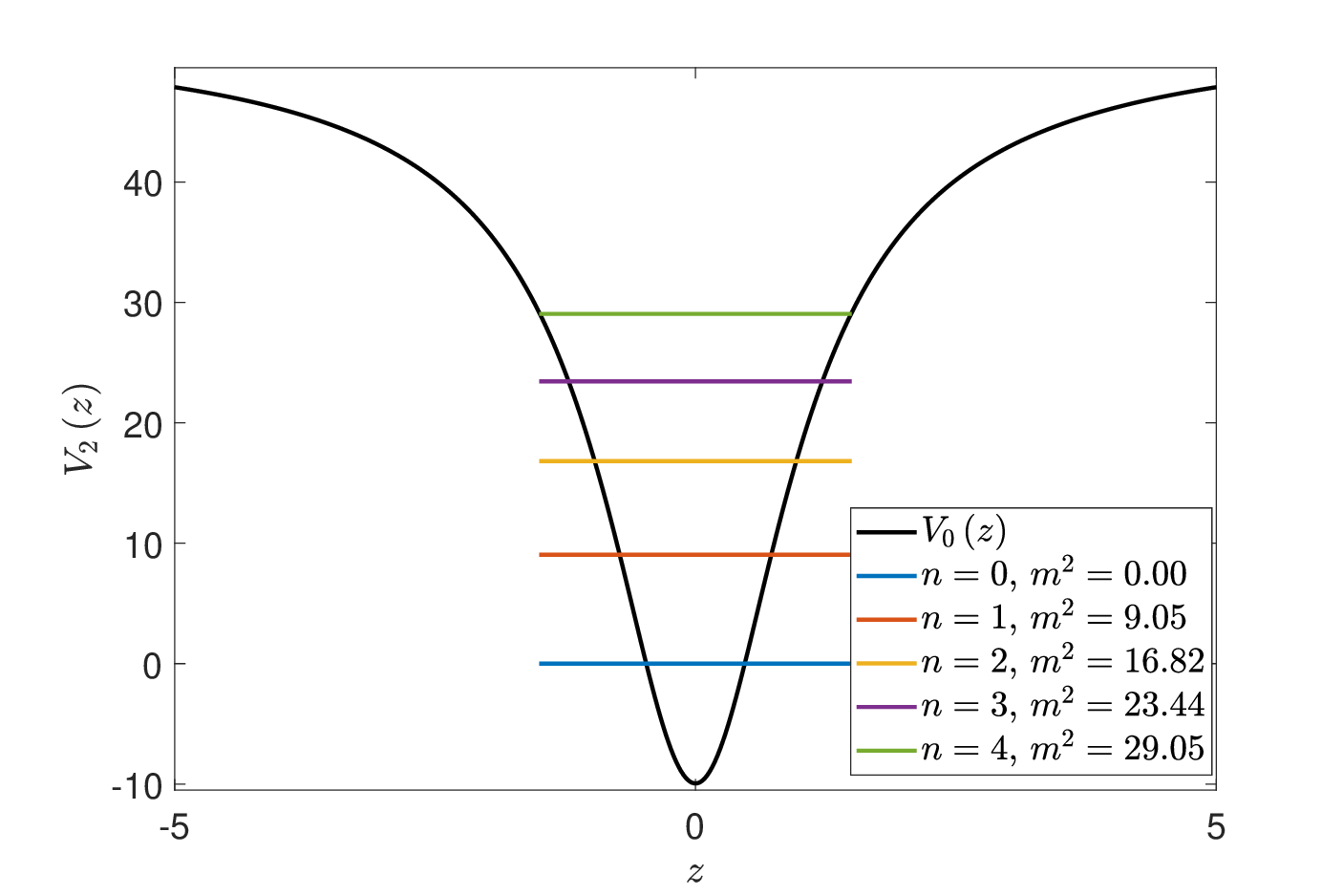}}
\subfigure[$\text C_1=10,k=1,\text C_2=0.5$]{\label{fig:subfig8b}
\includegraphics[width=0.31\textwidth]{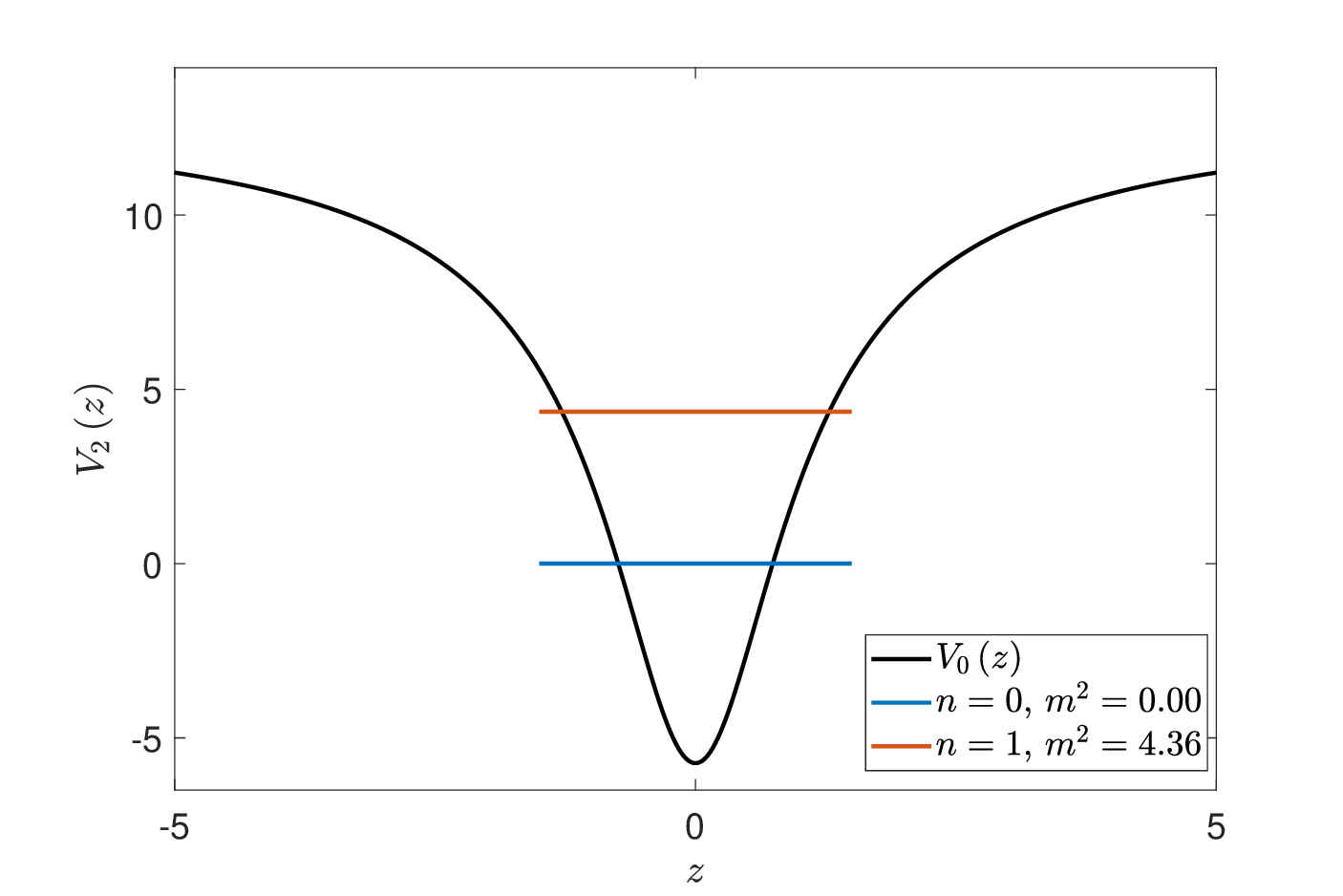}}
\subfigure[$\text C_1=10,k=1,\text C_2=0.6$]{\label{fig:subfig8c}
\includegraphics[width=0.31\textwidth]{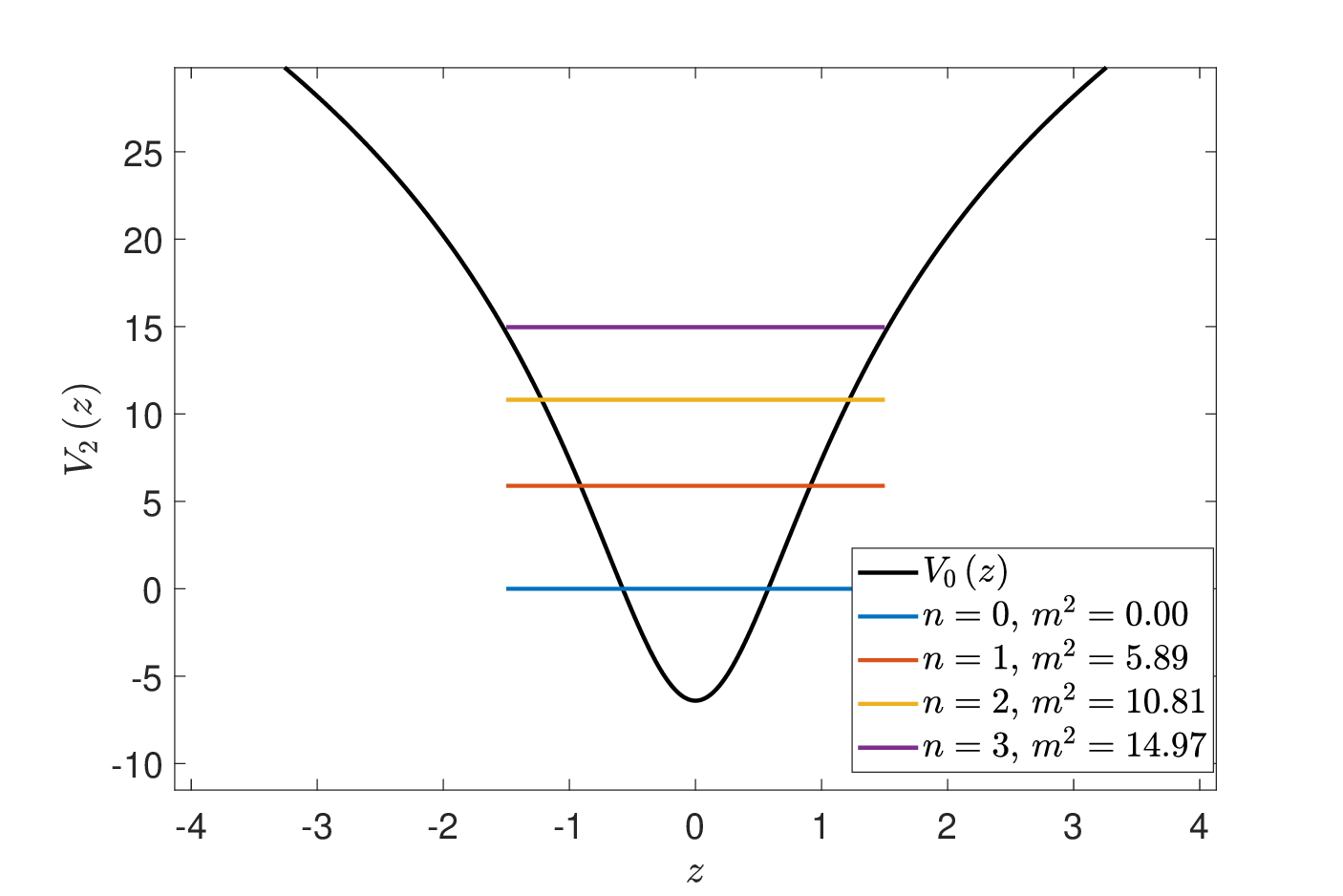}}
\subfigure[$\text C_1=20,k=1,\text C_2=0.5$]{\label{fig:subfig8d}
\includegraphics[width=0.31\textwidth]{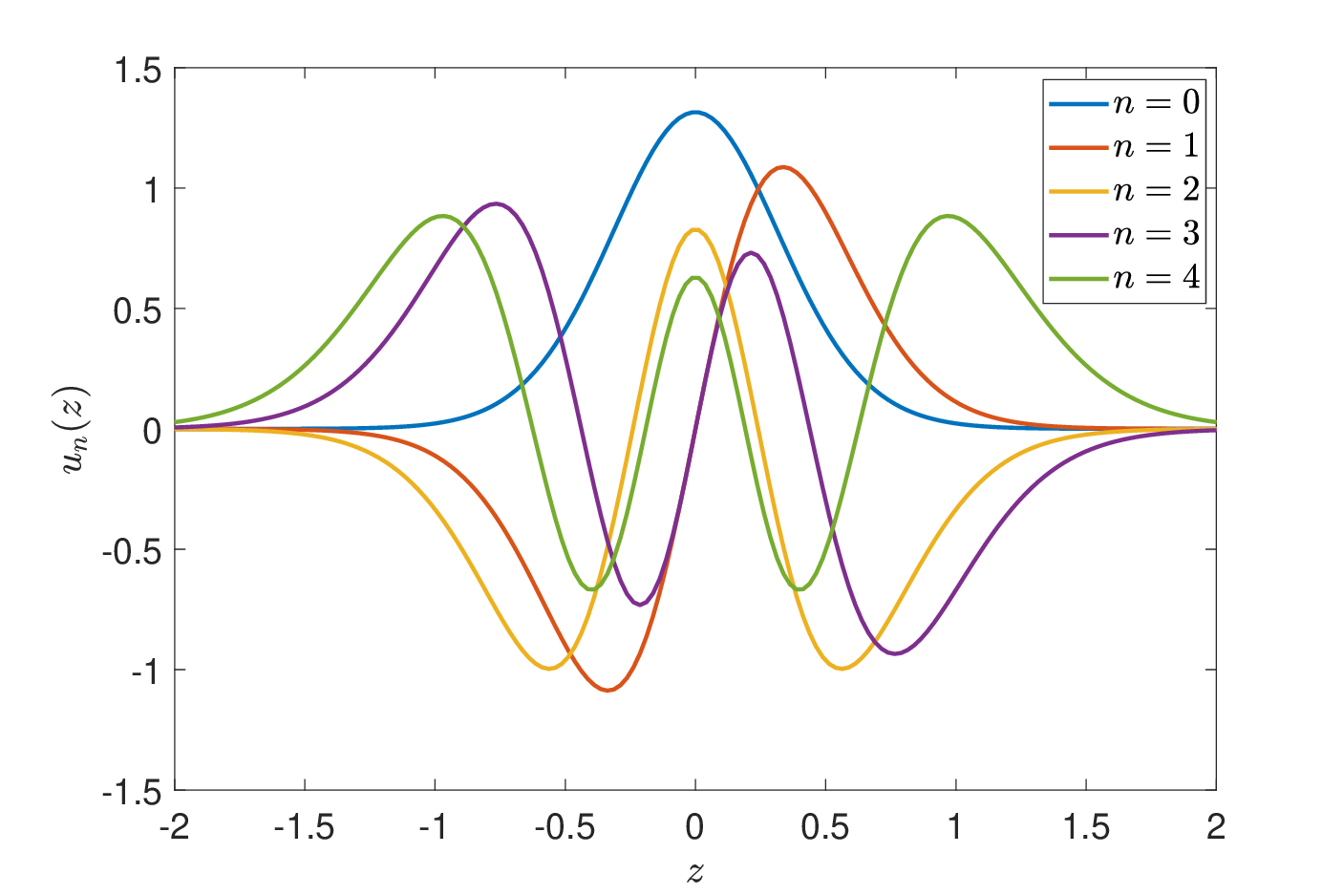}}
\hspace{0.1cm}
\subfigure[$\text C_1=10,k=1,\text C_2=0.5$]{\label{fig:subfig8e}
\includegraphics[width=0.31\textwidth]{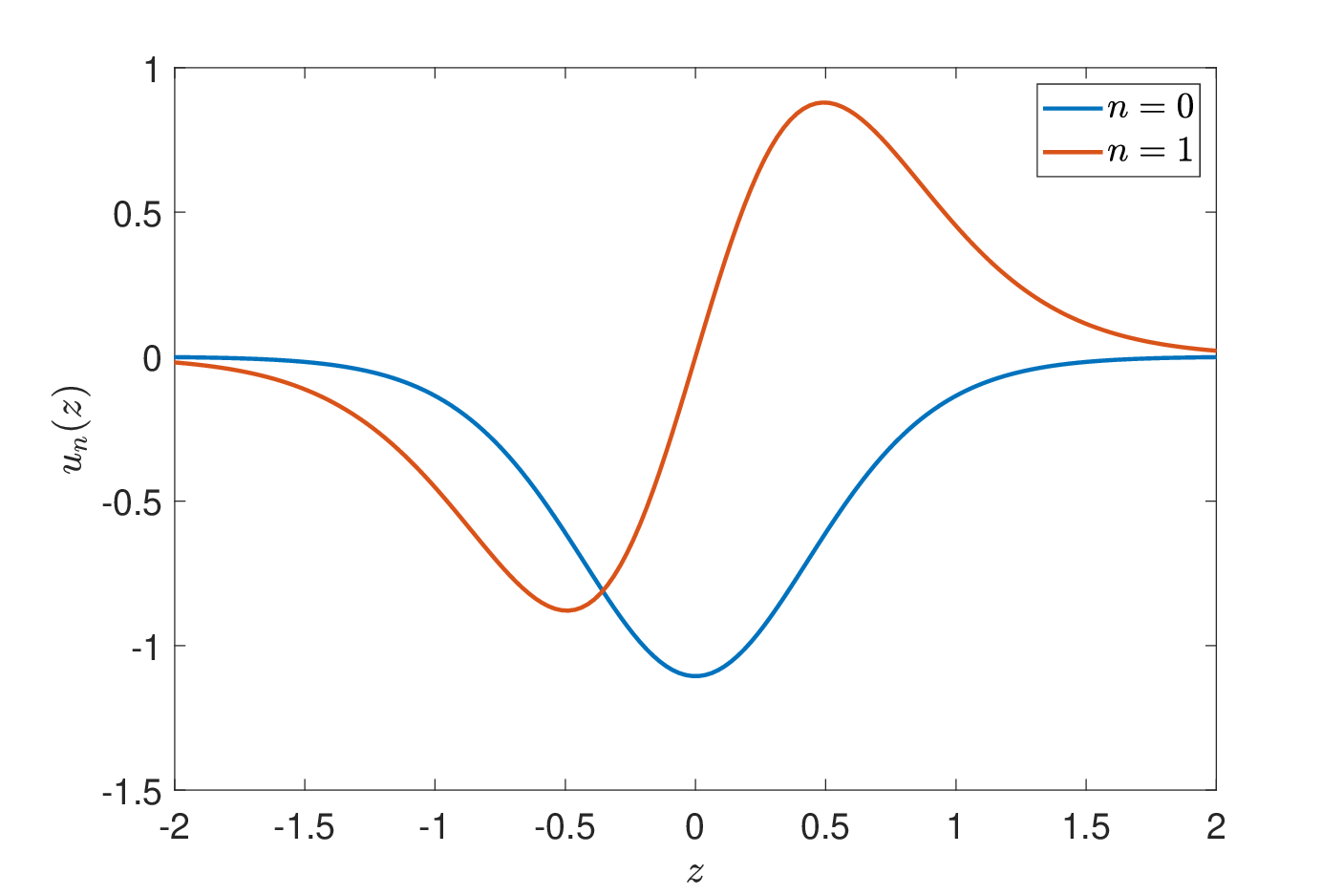}}
\hspace{0.1cm}
\subfigure[$\text C_1=10,k=1,\text C_2=0.6$]{\label{fig:subfig8f}
\includegraphics[width=0.31\textwidth]{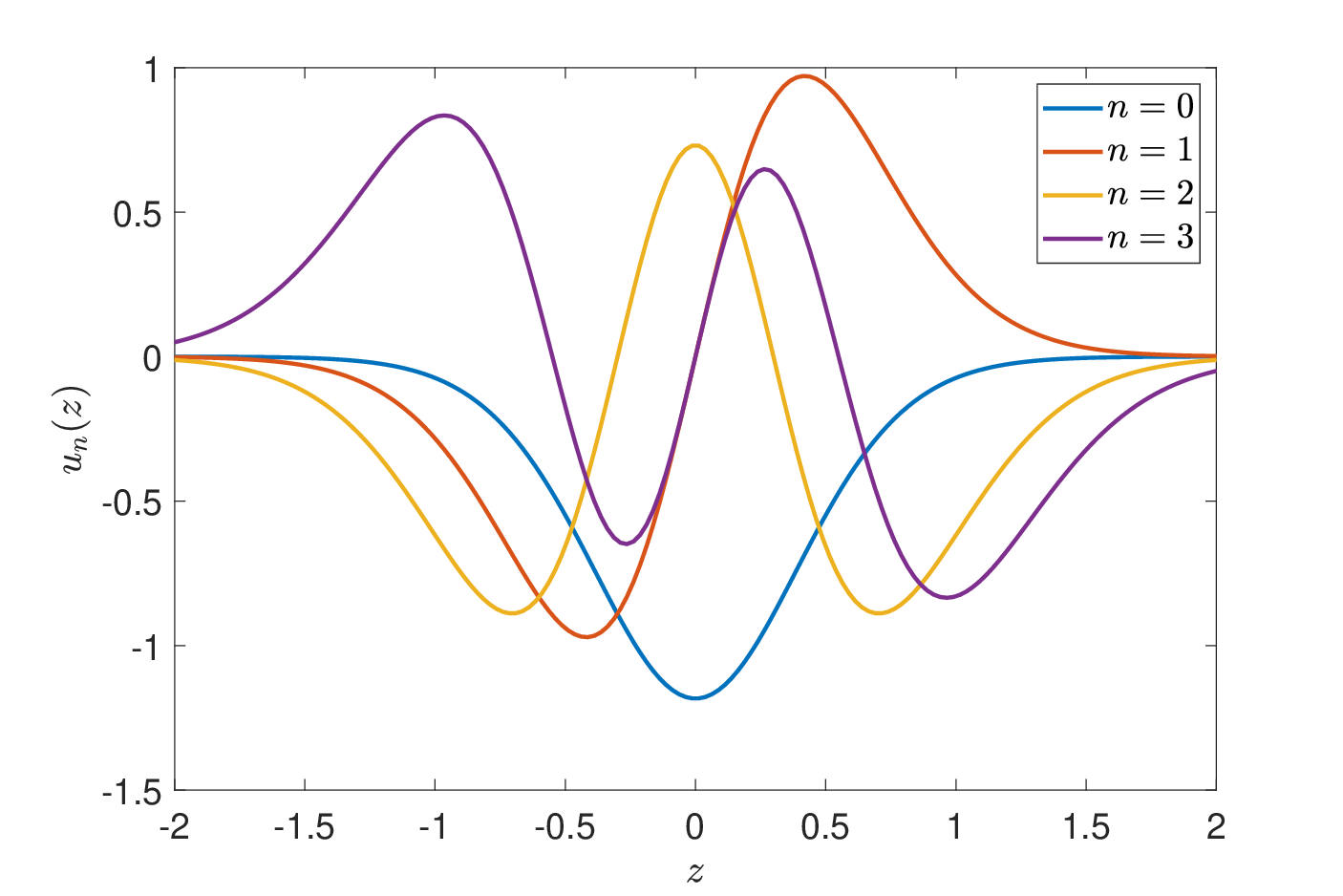}}
\caption{The upper row shows the potential $\displaystyle V_2\left(z\right)$ with the
         black line representing the potential and the colored lines indicating the
         position of mass spectra. The lower row presents the corresponding solutions
         of $\displaystyle u_n\left(z\right)$. Parameters are varied as
         $\text C_1=20,10,10; k=1$ and $\text C_2=0.5,0.5,0.6$, respectively.}
\label{fig:subfig_8}
\end{figure*}

To illustrate the above analysis, we plot the effective potential and the zero-mode in Fig. \ref{fig:subfig_7}.
The Fig. \ref{fig:subfig7a} demonstrates that the behaviors of the effective potential align well with the
conclusions (\ref{eq83}). Additionally for the case of $\text C_2=1/2$, a number of bound states are computed and
plotted in Fig. \ref{fig:subfig_8}, with their wave functions and the effective potential.

\begin{figure}[htbp]
\subfigure[$V_\varphi(z)$]{\label{fig:subfig9a}
\includegraphics[width=0.95\columnwidth]{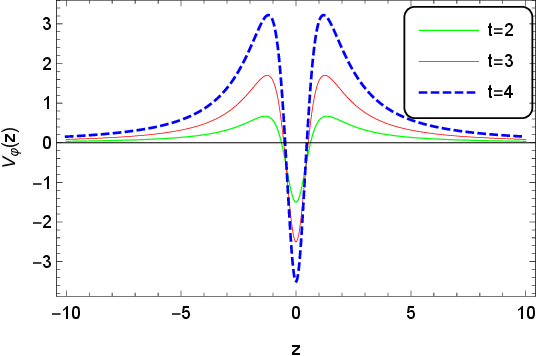}}
\subfigure[$\chi_\varphi(z)$]{\label{fig:subfig9b}
\includegraphics[width=0.95\columnwidth]{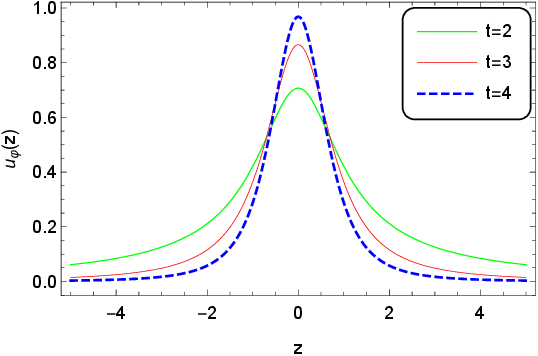}}
\caption{The effective potentials $V_{\varphi}(z)$ in (a), and the zero-mode of the KR field $u_\varphi(z)$
         in (b). The parameters are set as $\lambda=10, k=1$ and $t=2, 3, 4$.}
\label{fig:subfig_9}
\end{figure}

$Case$ II:
Coupling with the background scalar field, $\displaystyle F\left(R,\varphi\right)=\tilde F(\varphi)$.
Substituting equations (\ref{eq29}) and (\ref{eq30}) into the expressions for the effective potential (\ref{eq74})
and the zero-mode (\ref{eq77}), we obtain the zero-mode function
\begin{equation} \label{eq85}
u_{\varphi}(z) =N_2\left(\frac{147}{29}\right)^{\frac12t}\lambda^{\frac12} (1+k^2z^2)^{-\frac12t+\frac{1}{4}},
\end{equation}
and the effective potential
\begin{equation} \label{eq86}
V_{\varphi}(z) =\frac{k^2[2-4t+k^2(4t^2-1)z^2]}{4(1+k^2z^2)^2}.
\end{equation}
As $z\rightarrow\infty$, we can deduce
\begin{eqnarray} \label{eq87}
  u_{\varphi}(z\rightarrow\infty) &\propto& \frac{1}{z^{t-\frac{1}{2}}},     \\
  \quad V_{\varphi}(z\rightarrow\infty) &=& 0.
\end{eqnarray}
When \(t>{3}/{2}\), the zero-mode of the KR field is normalizable, and can be localized on the brane.
By selecting different values of \(t\), the effective potential and zero-mode can be obtained, as
depicted in Fig. \ref{fig:subfig_9}. Additionally, it can be observed that larger value of the
parameter \(t\) will lead to higher potential barrier.

In summary, with coupling to the gravity and the background scalar field, the zero-mode can be localized
on the brane, and the massive modes can also be localized when with certain values of the coupling
parameters. Specifically, the parameter \(\text C_2\) significantly impacts the convergence of the zero mode
when coupled to the gravity, and gives rise to varied behaviors of the effective potential at infinity.
The parameters \(\text C_1\) and \(k\) influence the depth of the potential well when the effective potential
approaches a finite value (as \(\text C_2=1/2)\), where larger values lead to deeper potential wells and a
larger number of the bound states. Conversely, the zero-mode of the KR field can be localized on the
brane when parameter $t>3/2$. Additionally, the parameter \(\lambda\) influences the zero-mode function.
The effective potential will always be a volcanic potential, with its barrier higher alongside the
parameter $t$.

\section{Conclusions}  \label{Sec55}
In this paper, we embark on a comprehensive examination of the localization phenomena on the domain
wall brane within the context of squared curvature gravity in a 5D spacetime framework. The initial
segment of our study is dedicated to reviewing the mechanisms through which the tensor field achieves
localization when coupled with the gravity in this brane model. Building upon this foundational
understanding, we extend our exploration to the localization dynamics of the $q$-form field on the
brane, taking into account its coupling interactions. The coupling function under consideration is
bifurcated into two distinct segments: one associated with the gravity, and the other with the
background scalar field. Through meticulous calculations, we derive the zero-mode solution, formulate
the Schr\"{o}dinger-like equation, and quantify the 4D effective action of the $q$-form field
localization on the brane. Furthermore, we delve into a detailed analysis of the localization
properties of the 0-form, 1-form, and 2-form fields when coupled to the gravitaty and the background
scalar field on a specific thick brane. Our investigation is enriched by numerical calculations,
from which we extract a diverse array of results depicted through relevant figures.

Our findings illuminate the critical role of the parameter $\text C_2$ in the localization
process of the $q$-form fields coupled to the gravity within a 5D spacetime. The localization behaviors
of the massive modes, described by the Schr\"{o}dinger-like equation at infinity, diverge into three
distinct scenarios contingent upon the values of $\text C_2$:
\begin{itemize}
\item For $\text C_2<1/2$, there is no localized massive mode;
\item For $\text C_2=1/2$, there will be a finite number of localized massive modes;
\item When $\text C_2>1/2$, an infinite discrete mass spectra emerges.
\end{itemize}
This delineation necessitates a discussion on the conditions where $\text C_2$ surpasses zero,
a prerequisite for the existence of localized zero modes.

Then, the coupling parameter $\displaystyle t$ emerges as a pivotal factor in the context of $q$-form field
coupling to the background scalar field exclusively. We suggest the positive value of the parameter $t$,
and the effective potential will always be in volcano form. Furthermore, for the 0-form fields, the
zero-mode can always be localized on the brane regardless of what value the parameter $t$ is. Conversely,
the existence of a localized zero-mode for the 1-form fields requires $\displaystyle t>{1}/{2}$, while
for 2-form fields, this threshold is raised to $\displaystyle t>{3}/{2}$.

{
Starting from the squared curvature gravity domain wall brane, we investigate the coupling of $q$-form
fields with gravity and the background scalar, and demonstrate that various $q$-form fields can be
localized on the thick brane. These scenarios further indicate the possible signatures arising from the
higher-dimensional spacetime, which could lead to observable features \cite{Capozziello2206.03690}. This
work thus provides a framework for exploring new signatures of higher-dimensional physics and motivates
further investigation into the phenomenology of bulk-brane systems.
}

In conclusion, our work not only addresses and fills the gaps left by previous studies that overlooked
scalar field coupling but also advances the analytical framework for examining the localization of
$q$-form fields on domain wall brane in the presence of the gravity and the background scalar field
couplings. By shedding light on these complex interactions, we pave the way for future investigations
into the diverse mechanisms of localization across different coupling forms, thereby enriching the
theoretical landscape of higher-dimensional physics and its implications for the fabric of our universe.

\section{acknowledgments}
The authors are extremely grateful for the anonymous referee, whose comments led to the improvement
of this paper.
This work was supported by the National Natural Science Foundation of China (Grants No. 11305119,
No. 11705070, and No. 11405121), the Natural Science Basic Research Plan in Shaanxi Province of
China (Program No. 2020JM-198), the Fundamental Research Funds for the Central Universities (Grants
No. JB170502), and the 111 Project (B17035).

\end{document}